\documentclass[a4paper,fleqn]{cas-sc}

\ExplSyntaxOn
\cs_if_exist:NF \msg_term:n { \cs_new_protected:Npn \msg_term:n #1 { \iow_term:n {#1} } }
\cs_set:Npn \__first_footerline: { \group_begin: \small \sffamily \__short_authors: \group_end: }
\ExplSyntaxOff

\usepackage{subfigure}
\usepackage{graphicx}
\usepackage{booktabs}
\newcolumntype{P}[1]{>{\raggedright\arraybackslash}p{#1}}
\usepackage{enumitem}
\usepackage{float}
\usepackage[section]{placeins}


\usepackage{amsthm}
\usepackage[numbers]{natbib}

\usepackage{algorithm}
\usepackage{algorithmic}
\usepackage[english]{babel}
\newtheorem{theorem}{Theorem}
\newtheorem{proposition}[theorem]{Proposition}
\newtheorem{lemma}[theorem]{Lemma}
\newtheorem{corollary}[theorem]{Corollary}
\theoremstyle{definition}
\newtheorem{definition}[theorem]{Definition}
\newtheorem{remark}[theorem]{Remark}

\newcommand{\R}{\mathbb{R}}
\newcommand{\C}{\mathbb{C}}
\newcommand{\T}{\mathbb{T}}
\newcommand{\G}{\mathcal{G}}
\newcommand{\Fo}{\mathcal{F}}
\newcommand{\re}{\operatorname{Re}}
\newcommand{\rank}{\operatorname{rank}}
\newcommand{\Lip}{\operatorname{Lip}}
\newcommand{\ii}{\mathrm{i}}
\newcommand{\dd}{\,\mathrm{d}}
\newcommand{\one}{\mathbf{1}}
\newcommand{\Tk}{\mathcal{T}_{\kappa}}
\newcommand{\Hk}{\mathcal{H}}
\newcommand{\sci}[2]{#1\times10^{#2}}

\newcommand{\hnorepo}{\href{https://github.com/BiswanathB/Hankel-Neural-Operator-}{\textbf{https://github.com/\allowbreak BiswanathB/\allowbreak Hankel-Neural-Operator-}}}

\def\tsc#1{\csdef{#1}{\textsc{\lowercase{#1}}\xspace}}
\tsc{WGM}
\tsc{QE}
\tsc{EP}
\tsc{PMS}
\tsc{BEC}
\tsc{DE}
\begin{document}
	\let\WriteBookmarks\relax
	\def\floatpagepagefraction{1}
	\def\textpagefraction{.001}
	\shorttitle{Hankel Neural Operator: A New Alternative for Operator Learning}
	\shortauthors{Barman, Ray}

	\title [mode = title]{Beyond Fourier and Laplace: The Hankel Neural Operator for Learning Solution Operators of Differential Equations}

	\author[1]{Biswanath Barman}
	\ead{biswa.nit.rourkela@gmail.com}

	\author[1]{Rajendra K. Ray$^{*}$}
	\cortext[cor1]{Corresponding author}
	\ead{rajendra@iitmandi.ac.in}

	\address[1]{School of Mathematical and Statistical Sciences, Indian Institute of Technology Mandi\\
	Mandi, Himachal Pradesh, 175005, India}

\begin{abstract}
Fourier and Laplace neural operators (FNO, LNO) learn kernel integral operators in a transform domain evaluated on a periodic grid. FNO has no explicit mechanism for boundaries, initial states or transients; LNO adds pole--residue symbols but evaluates them on the discrete Fourier grid with unconstrained poles. FNO's error is large on undamped oscillators and grows under grid refinement. We introduce the Hankel neural operator (HNO), derived from the structure of the solution operator. We prove that the solution operator of a well-posed linear constant-coefficient boundary-value problem on an interval is a Toeplitz operator with a rational kernel plus a finite-rank Hankel-type operator carrying boundary information. Classical results of Kronecker, Nehari and Adamyan--Arov--Krein characterise this kernel class: stable poles, finite Hankel rank and norm bounds independent of the interval length. HNO realises the class in three ways: a non-periodic convolution evaluated by the FFT, with a causal form; a closed-loop Lur'e operator for dynamics without fading memory; and a rational Fourier multiplier that contains FNO as a special case. Trained from data alone, HNO has the lowest mean test error on all nine LNO benchmarks and on the Burgers, Darcy and Navier--Stokes benchmarks of FNO; two of these differences, and the Navier--Stokes comparison with a parameter-matched FNO, are not statistically significant. On the six ODE problems it lowers the better baseline's mean error 2.1--5.2 times, and its Darcy error stays nearly constant from $85^2$ to $421^2$ grids, where FNO's grows eightfold. Ablations support each component. Code and data: \hnorepo{}.
\end{abstract}

\begin{keywords}
Neural operators \sep Operator learning \sep Hankel operators \sep Toeplitz operators \sep Lur'e systems \sep Fading memory \sep Scientific machine learning \sep Partial differential equations
\end{keywords}

	\date{}
	\maketitle

%============================= INTRODUCTION ===============================
\section{Introduction}
\label{sec:intro}

Many tasks in computational science require the solution of the same differential equation for a large number of inputs: forcing histories in structural dynamics, coefficient fields in subsurface flow, or initial conditions in weather and climate models. Classical solvers treat each input as a new problem. Neural operators instead learn the solution operator $\G:a\mapsto u$ itself from input--output pairs, after which a new solution costs one forward pass \cite{lu2021deeponet,li2021fno,kovachki2023neural,azizzadenesheli2024neural}. The approach rests on universal approximation results for operators \cite{chen1995universal,kovachki2021universal,lanthaler2022error}, and its practical value depends on how well the architecture's inductive bias matches the operators met in practice \cite{boulle2024mathematical,lu2022comprehensive}.

The Fourier neural operator (FNO) \cite{li2021fno} composes layers of the form $v\mapsto\sigma(\Fo^{-1}[\Phi\,\Fo v]+Wv)$, where the learned multiplier $\Phi$ is kept for a few low Fourier modes. It is fast and accurate on periodic problems. Its kernel, however, is a band-limited function on a torus. A solution operator on a bounded interval or box is not translation invariant: it carries information from the boundaries, from the initial state and, for forced dynamics, from the past of the input. FNO represents none of these explicitly. Zero padding by a fixed number of grid points is the usual remedy, but it changes the operator when the grid changes. The Laplace neural operator (LNO) \cite{cao2024lno} identified this bottleneck and replaced the Fourier multiplier with a pole--residue symbol $\sum_n\beta_n/(s-\mu_n)$, which describes transient and steady responses with the same parameters. LNO improved markedly on FNO for undamped oscillators. Yet its steady part is still a circular convolution on the DFT grid, its transient coefficients are computed from the DFT of the input (a periodic extension), and its poles are unconstrained, so a layer may contain growing modes. On the LNO benchmarks themselves its seed-to-seed spread is large (for the Euler--Bernoulli beam the standard deviation exceeds the mean; Table~\ref{tab:lno}).

This paper starts from a different question: \emph{what does the solution operator of a differential equation on a bounded domain look like?} For linear constant-coefficient problems on an interval, linear systems theory gives an exact answer. The operator is a Toeplitz part, a convolution with a Green's function whose transform is rational, plus a finite-rank Hankel-type part that collects everything the boundaries contribute (Theorem~\ref{thm:structure}). Hankel operators are well understood. Kronecker's theorem equates finite rank with a rational symbol, Nehari's theorem bounds the norm, and the Adamyan--Arov--Krein (AAK) theorem gives the optimal low-rank approximation \cite{peller2003hankel,partington1988introduction,nehari1957bounded,adamjan1971analytic,glover1984all}. These results have long been used for model reduction and system identification \cite{ho1966effective,juang1985eigensystem,antoulas2005approximation}, but not as the design principle of a neural operator. We use them as such and call the result the \emph{Hankel neural operator} (HNO). The design question thereby changes from which integral transform to apply to which structure the solution operator has.

The main contributions are as follows.
\begin{enumerate}[leftmargin=*,itemsep=1pt]
\item A structural theorem (Theorem~\ref{thm:structure}) and its causal form (Theorem~\ref{thm:causal}) that identify the kernel class of linear solution operators on intervals, together with an HNO layer that represents this class exactly (Corollary~\ref{cor:exact}) and whose norm bounds, for fixed parameters, do not depend on the domain length (Proposition~\ref{prop:bounded}). On boxes the same class is used along each axis through tensor products.
\item A closed-loop (Lur'e) HNO for dynamics without fading memory. Its motivation is a quantitative fading-memory bound that we prove for feed-forward operators with stable kernels (Proposition~\ref{prop:wiener}), and it is evaluated by block time marching in which the linear state is propagated exactly (Proposition~\ref{prop:semigroup}).
\item A frequency-domain realisation on periodic and padded domains: a rational Fourier multiplier defined at every frequency, which contains FNO as the special case of zero residues (Proposition~\ref{prop:fno}) and has a resolution-independent parameterisation when the padding is a fixed fraction of the domain.
\item A complete evaluation on all nine problems of the LNO suite \cite{cao2024lno} and the three problems of the FNO suite \cite{li2021fno}, each under the protocol of the original paper, with significance tests, resolution transfer, cost, and ablations of every design choice. We also state where the gains are small or not statistically significant, and why.
\end{enumerate}

The remainder of the paper is organised as follows. Section~\ref{sec:related} reviews related work, Section~\ref{sec:prelim} fixes notation and recalls the facts on Hankel operators, Section~\ref{sec:method} develops the HNO, Section~\ref{sec:numerics} reports the numerical study, Section~\ref{sec:discussion} discusses advantages, limitations and a research plan, and Section~\ref{sec:conclusion} concludes.

%============================= RELATED WORKS ===============================
\section{Related Works}
\label{sec:related}

\textit{Neural operators.} DeepONet \cite{lu2021deeponet} realised the universal approximation theorem of Chen and Chen \cite{chen1995universal} with branch and trunk networks. The neural-operator framework of Kovachki et al. \cite{kovachki2023neural} composes learned integral operators with pointwise nonlinearities; FNO \cite{li2021fno} evaluates the integral operator by the FFT. Many variants modify the integral operator or the domain: wavelet and multiwavelet operators \cite{tripura2023wavelet,gupta2021multiwavelet}, U-shaped and factorised FNOs \cite{rahman2023uno,tran2023factorized,wen2022ufno}, learned deformations for general geometries \cite{li2023geofno}, spectral neural operators with fixed bases \cite{fanaskov2023spectral}, attention-based operators \cite{cao2021choose,hao2023gnot} and physics-informed training \cite{raissi2019physics,li2024pino,goswami2023physics}. Alias-free formulations \cite{bartolucci2023representation} address the dependence of neural operators on the discretisation. Comparisons under common data and protocols \cite{lu2022comprehensive} show that the ranking of architectures depends strongly on the problem class.

\textit{Pole--residue and state-space models.} LNO \cite{cao2024lno} is the closest work: it introduced trainable poles and residues into operator learning. In sequence modelling, structured state-space models \cite{gu2022efficiently} and linear recurrent units \cite{orvieto2023resurrecting} use diagonal linear dynamics with stable eigenvalues and convolution or scan evaluation. HNO shares the stable modal parameterisation with these models but differs in three respects: it is derived from the structure of solution operators on intervals, it includes the finite-rank Hankel (boundary) term that sequence models do not have, and it is closed through a static nonlinearity in the Lur'e form for non-fading-memory dynamics.

\textit{Green's functions and system identification.} Learning Green's functions directly \cite{gin2021deepgreen,melchers2024neural,yoo2025neural,boulle2024mathematical} targets linear operators with a low-rank structure, which is related to ours. Toeplitz matrices have also been used for token mixing in sequence models \cite{qin2023toeplitz}, without a boundary (Hankel) term or a link to solution operators. Hankel matrices are the classical tool for identifying a state-space model from input--output data \cite{ho1966effective,juang1985eigensystem}, rational approximation of transfer functions is done by vector fitting \cite{gustavsen1999rational}, and the Hankel singular values certify model reduction \cite{glover1984all,antoulas2005approximation,zhou1996robust}. Dynamic mode decomposition \cite{schmid2010dynamic} and Koopman methods likewise represent dynamics by modes. Boyd and Chua \cite{boyd1985fading} showed that causal operators with fading memory are approximated by stable linear systems followed by a static map; the converse limitation motivates our closed-loop construction. To our knowledge, no previous neural operator parameterises its layers by the Toeplitz plus finite-rank Hankel-type decomposition of solution operators.

%============================= PRELIMINARY ===============================
\section{Preliminaries}
\label{sec:prelim}

\subsection{Operator learning and notation}

Let $\Omega\subset\R^d$ be a bounded interval or box, or a torus $\T^d$, and let $\mathcal A$ and $\mathcal U$ be Banach spaces of functions on $\Omega$. Given $N_{\rm tr}$ pairs $\{(a_j,\G(a_j))\}$ of an unknown operator $\G:\mathcal A\to\mathcal U$, a neural operator $\G_\theta$ is trained by minimising the empirical relative $L^2$ risk
\begin{equation}
\min_\theta\ \frac{1}{N_{\rm tr}}\sum_{j=1}^{N_{\rm tr}}\frac{\|\G_\theta(a_j)-\G(a_j)\|_{L^2(\Omega)}}{\|\G(a_j)\|_{L^2(\Omega)}} .
\label{eq:risk}
\end{equation}
FNO, LNO and HNO share the architecture \cite{kovachki2023neural}
\begin{equation}
\G_\theta=\mathcal Q\circ\mathcal L_L\circ\cdots\circ\mathcal L_1\circ\mathcal P,\qquad
\mathcal L_\ell(v)=\sigma\big(\mathcal K_\ell v+W_\ell v+b_\ell\big),
\label{eq:template}
\end{equation}
where $\mathcal P$ lifts the input pointwise to $C$ channels, $\mathcal Q$ projects to the output, $W_\ell\in\R^{C\times C}$ is a pointwise linear map, $\sigma$ is a pointwise nonlinearity and $\mathcal K_\ell$ is a linear integral operator with kernel $\kappa$, $(\mathcal K v)_o(x)=\sum_i\int_\Omega\kappa_{oi}(x,y)v_i(y)\dd y$. The three operators differ only in the kernel class and in how $\mathcal K$ is evaluated. Table~\ref{tab:notation} lists the notation used throughout.

\begin{table}[pos=htbp]
\centering\small
\caption{Notation.}
\label{tab:notation}
\begin{tabular}{@{}ll@{\qquad}ll@{}}
\toprule
$a,\ u$ & input and output functions & $\lambda=-\eta+\ii\omega$ & pole, decay rate $\eta=e^{\alpha}>0$\\
$\G,\ \G_\theta$ & true and learned operator & $R^{\pm}$ & causal / anti-causal residues\\
$v$, $C$, $L$ & hidden function, width, depth & $N$ & poles per channel\\
$\kappa,\ \hat\kappa$ & kernel and its symbol & $\zeta^{(0)},\zeta^{(1)}$ & boundary states\\
$\Tk$ & Toeplitz (convolution) operator & $\mathbf M$, $G$ & Hankel mixing, gains\\
$\Gamma_\kappa$, $\Hk$ & Hankel operator, Hankel-type part of a layer & $W,\ D$ & local map, direct feedthrough\\
$\Fo$, $\xi$ & Fourier transform, frequency & $\Phi(k)$, $k_{\max}$ & free Fourier modes\\
$n$ & grid points per axis & $\phi_n(\xi)$ & normalised rational term\\
$z,\ e,\ y$ & Lur'e state, error signal, output & $F$ & static feedback\\
$J,\ \Delta_b,\ q$ & time blocks, block length, sweeps & $\varepsilon$ & padding fraction\\
$s_j$ & Hankel singular values & $\Pi$ & Hermitian projection\\
\bottomrule
\end{tabular}
\end{table}

\subsection{The two baselines and where they fail}
\label{sec:baselines}

\textit{FNO} \cite{li2021fno}. The domain is treated as a torus and $\mathcal K v=\Fo^{-1}[\Phi(k)\,\Fo v]$ with free complex weights $\Phi(k)\in\C^{C\times C}$ for $|k|<k_{\max}$ and $\Phi(k)=0$ otherwise. The kernel is a band-limited periodic function. Non-periodic problems are handled by padding a fixed number of grid points.

\textit{LNO} \cite{cao2024lno}. The symbol is $\hat\kappa_{oi}(s)=\sum_n\beta_{oin}/(s-\mu_{oin})$. For an input with DFT coefficients $\alpha_l$ at $\ii\omega_l$, the response is split into a steady part $\sum_l\alpha_l\hat\kappa(\ii\omega_l)e^{\ii\omega_lt}$, a circular convolution, and a transient part $\sum_n\gamma_ne^{\mu_nt}$ with $\gamma_n=\beta_n\sum_l\alpha_l/(\mu_n-\ii\omega_l)$. The poles $\mu_n$ are unconstrained. The single-layer LNO is used for the LNO suite; for the FNO suite we use the multi-layer LNO of the official repository, which adds a separate Fourier-mode branch.

Two structural gaps follow. First, neither model has an explicit representation of the boundary or initial-state contribution to the solution: in FNO it can only be imitated by $W$ and padding, and in LNO it is estimated from a periodic extension of the input. Second, both are feed-forward compositions without a state carried across the window, so information from early forcing reaches late times only through long, slowly decaying kernels (Section~\ref{sec:lure}). The undamped oscillators of the LNO suite need exactly this, because the test forcing decays and the late response is the free oscillation launched early in the window. FNO's test error there is 0.70--0.79 (Table~\ref{tab:lno}).

\subsection{Hankel operators}
\label{sec:hankel}

For a kernel $\kappa\in L^1(\R_+)$ the Hankel operator maps the past of an input to the future of the output,
\begin{equation}
(\Gamma_\kappa f)(t)=\int_0^\infty\kappa(t+s)\,f(s)\dd s,\qquad t\ge0 .
\label{eq:hankel}
\end{equation}
After the substitution $f(s)\mapsto f(-s)$, $\Gamma_\kappa$ is the part of $\kappa*f$ that a causal system still remembers at $t=0$ from inputs applied before $t=0$. In discrete time it is the Hankel matrix $[\kappa_{j+l}]_{j,l\ge0}$. We use three classical facts \cite{peller2003hankel,partington1988introduction}.

\begin{theorem}[Kronecker \cite{peller2003hankel,partington1988introduction}]\label{thm:kronecker}
For $\kappa\in L^1(\R_+)$, $\rank\Gamma_\kappa\le N$ if and only if its Laplace transform $\hat\kappa$ is a strictly proper rational function of McMillan degree at most $N$. For distinct poles ($\re\lambda_n<0$), $\kappa(t)=\sum_{n=1}^NR_ne^{\lambda_nt}$ and $\Gamma_\kappa=\Psi_o\Psi_c$ factors through $\C^N$, with $\Psi_cf=(\int_0^\infty e^{\lambda_ns}f(s)\dd s)_n$ (the state at $t=0$) and $\Psi_oz=\sum_nR_nz_ne^{\lambda_nt}$.
\end{theorem}

\begin{theorem}[Nehari \cite{nehari1957bounded,partington1988introduction}]\label{thm:nehari}
$\|\Gamma_\kappa\|_{L^2\to L^2}$ equals the $L^\infty$ distance of the symbol $\hat\kappa(\ii\xi)$ to the boundary values of bounded analytic functions in the left half-plane; in particular $\|\Gamma_\kappa\|\le\|\hat\kappa\|_{L^\infty}$. For $\kappa(t)=\sum_nR_ne^{\lambda_nt}\one_{t\ge0}$ with $\re\lambda_n<0$, both the convolution and the Hankel operator are bounded on $L^2$, and $\|\kappa*\cdot\|\le\|\kappa\|_{L^1}\le\sum_n|R_n|/|\re\lambda_n|$.
\end{theorem}

\begin{theorem}[Adamyan--Arov--Krein \cite{adamjan1971analytic,glover1984all}]\label{thm:aak}
Let $s_1\ge s_2\ge\cdots$ be the singular values of $\Gamma_\kappa$. Among all kernels whose Hankel operator has rank at most $r$, the smallest error $\|\Gamma_\kappa-\Gamma_{\tilde\kappa}\|$ is $s_{r+1}$. Balanced truncation to order $r$ satisfies $\|\hat\kappa-\hat{\tilde\kappa}\|_\infty\le2\sum_{j>r}s_j$.
\end{theorem}

For neural operators these results mean that (i) a complex pole--residue kernel with $N$ distinct poles and non-zero residues has Hankel rank exactly $N$ (at most $2N$ for its real part), (ii) the constraint $\re\lambda<0$ gives bounds that do not depend on the length of the domain, and (iii) for a single linear layer, the number of poles is a model-order choice with a computable error certificate. Theorem~\ref{thm:aak} bounds the approximation error of one linear kernel; it does not by itself certify the error of the full nonlinear network.

%============================= METHODOLOGY ===============================
\section{Methodology: the Hankel Neural Operator}
\label{sec:method}

\subsection{Structure of solution operators on an interval}
\label{sec:structure}

Let $\mathsf L=\sum_{k=0}^{m}c_k\,\mathrm d^k/\mathrm dt^k$ with $c_m=1$ have characteristic polynomial $p(s)=\sum_kc_ks^k$ with distinct roots $\lambda_1,\dots,\lambda_m$, none on the imaginary axis. Impose $m$ linear homogeneous conditions $\mathcal B_ju=0$, each a combination of $u^{(r)}(0)$ and $u^{(r)}(1)$, $r\le m-1$, such that $\mathsf Lu=f$, $\mathcal Bu=0$ has a unique solution $u=\G f$ for every $f\in L^2(0,1)$.

\begin{theorem}[Toeplitz plus finite-rank Hankel-type term]\label{thm:structure}
Let $\Lambda_\pm=\{n:\re\lambda_n\gtrless0\}$ and $r_n=1/p'(\lambda_n)$. Then
\begin{equation}
(\G f)(t)=\underbrace{\int_0^1\kappa(t-s)f(s)\dd s}_{(\Tk f)(t)}
+\underbrace{\sum_{n\in\Lambda_-}c_n(f)\,e^{\lambda_nt}+\sum_{n\in\Lambda_+}c_n(f)\,e^{-\lambda_n(1-t)}}_{(\Hk f)(t)},
\label{eq:structure}
\end{equation}
with the decaying free-space Green's function
\begin{equation}
\kappa(\tau)=\sum_{n\in\Lambda_-}r_ne^{\lambda_n\tau}\one_{\tau\ge0}-\sum_{n\in\Lambda_+}r_ne^{\lambda_n\tau}\one_{\tau<0}.
\label{eq:green}
\end{equation}
The coefficients are $c=-\mathbf M^{-1}\mathcal B[\Tk f]$ for an invertible $\mathbf M\in\C^{m\times m}$, and $\mathcal B[\Tk f]$ is a linear function of the $2m$ boundary states
\begin{equation}
\zeta^{(1)}_n(f)=\int_0^1e^{\mu_n(1-s)}f(s)\dd s,\qquad
\zeta^{(0)}_n(f)=\int_0^1e^{\mu_ns}f(s)\dd s,\qquad
\mu_n=\begin{cases}\lambda_n,&n\in\Lambda_-,\\-\lambda_n,&n\in\Lambda_+,\end{cases}\quad\re\mu_n<0 .
\label{eq:bstates}
\end{equation}
Hence $\Hk=-\Psi_o\mathbf M^{-1}\widetilde{\mathcal B}\Psi_c$, with $\Psi_c$ the map to the boundary states and $\widetilde{\mathcal B}$ a fixed matrix, has rank at most $m$ (its range is spanned by $m$ exponentials), all exponentials in \eqref{eq:structure}--\eqref{eq:bstates} decay, and $\G$ is fixed by $m$ stable exponents, the residues and $\mathbf M$.
\end{theorem}

\begin{proof}
Partial fractions give $1/p(s)=\sum_nr_n/(s-\lambda_n)$. The two-sided inverse Laplace transform in the strip that contains the imaginary axis is \eqref{eq:green}, so $\mathsf L\kappa=\delta$ and $\mathsf L(\Tk f)=f$ on $(0,1)$. Therefore $u_h=\G f-\Tk f$ satisfies $\mathsf Lu_h=0$ and lies in the span of $e^{\lambda_nt}$. For $n\in\Lambda_+$ we write $e^{\lambda_nt}=e^{\lambda_n}e^{-\lambda_n(1-t)}$, a decaying exponential anchored at $t=1$, which gives the form of $\Hk$. The conditions $\mathcal B(\Tk f+u_h)=0$ read $\mathbf Mc=-\mathcal B[\Tk f]$, where $\mathbf M_{jn}$ is $\mathcal B_j$ applied to the $n$-th basis function; $\mathbf M$ is invertible because the homogeneous problem has only the trivial solution. Finally, $(\Tk f)^{(r)}(\theta)=\int_0^1\kappa^{(r)}(\theta-s)f(s)\dd s$ for $\theta\in\{0,1\}$ and $r\le m-1$, and $\kappa^{(r)}(\theta-s)$ is a combination of $e^{\mu_n(1-s)}$ (for $\theta=1$) and $e^{\mu_ns}$ (for $\theta=0$), i.e.\ of the boundary states. Only $m$ of the $2m$ states are needed ($\zeta^{(1)}_n$ for $n\in\Lambda_-$ and $\zeta^{(0)}_n$ for $n\in\Lambda_+$); the layer below uses all of them for simplicity.
\end{proof}

We call $\Hk$ a \emph{Hankel-type} term because, as in Theorem~\ref{thm:kronecker}, it factors through a finite-dimensional state: the input enters only through $\Psi_c$ and the output is a combination of exponentials ($\Psi_o$). It is not the operator $\Gamma_\kappa$ of \eqref{eq:hankel} itself, but it reduces to $\Gamma_\kappa$ for forcing applied before the window (Remark~\ref{rem:past}).

\begin{remark}\label{rem:past}
Repeated roots produce $t^ke^{\lambda t}$, which are limits of distinct poles, and purely imaginary roots (undamped oscillators) are limits $\re\mu\to0^-$ that the stable parameterisation approximates arbitrarily well on a bounded interval, at the price of large constants in Proposition~\ref{prop:bounded}. For a causal kernel and the bi-infinite convolution, $\int_{-\infty}^t\kappa(t-s)f(s)\dd s=(\Tk f)(t)+(\Gamma_\kappa f_{\rm past})(t)$ with $f_{\rm past}(s)=f(-s)$, $s>0$: forcing applied before the observation window and non-zero initial states enter exactly through a finite-rank Hankel-type term. A periodic convolution has no such term.
\end{remark}

\begin{theorem}[Initial-value problems are causal]\label{thm:causal}
If all $m$ conditions are imposed at $t=0$, $u^{(r)}(0)=u_r$, then
\begin{equation}
(\G f)(t)=\int_0^tg(t-s)f(s)\dd s+\sum_nc_n(u_0,\dots,u_{m-1})\,e^{\lambda_nt},\qquad g(\tau)=\sum_nr_ne^{\lambda_n\tau}.
\label{eq:causal}
\end{equation}
The Hankel-type term depends only on the initial data and vanishes when they are zero.
\end{theorem}

\begin{proof}
The causal impulse response satisfies $\mathsf Lg=\delta$ with $g^{(r)}(0^+)=0$ for $r<m-1$, so the Volterra integral solves $\mathsf Lu=f$ with zero initial data; the remainder solves $\mathsf Lu_h=0$ with the prescribed data. The output at $t$ uses $f|_{[0,t]}$ only.
\end{proof}

Theorems~\ref{thm:structure} and~\ref{thm:causal} fix the kernel class that a linear operator layer should represent on an interval: a rational Toeplitz kernel with stable exponents, a finite-rank Hankel-type term driven by boundary states, and, on time axes, causality. Neither a band-limited periodic kernel (FNO) nor a pole--residue symbol evaluated on a periodic grid (LNO) belongs to this class. For boxes we use tensor products of this class along each axis, and for nonlinear operators we compose such layers with pointwise maps. These two extensions are motivated by the theorems but not covered by them; their justification is empirical (Section~\ref{sec:numerics}).

\subsection{The Toeplitz--Hankel HNO layer}
\label{sec:layer}

\begin{definition}[Stable pole bank]
For each input channel $i$ and $n=1,\dots,N$, $\lambda_{in}=-e^{\alpha_{in}}+\ii\,\omega_{in}$ with free parameters $\alpha_{in},\omega_{in}\in\R$. Thus $\re\lambda_{in}<0$ for all parameter values.
\end{definition}

\begin{definition}[Toeplitz--Hankel HNO layer]\label{def:layer}
On an interval mapped to $[0,1]$ and for $v:[0,1]\to\R^C$,
\begin{align}
\kappa_{oi}(\tau)&=\re\sum_nR^{+}_{oin}e^{\lambda_{in}\tau}\ (\tau\ge0),\qquad
\kappa_{oi}(\tau)=\re\sum_nR^{-}_{oin}e^{\lambda_{in}|\tau|}\ (\tau<0),\label{eq:hk}\\
\zeta_{in}&=\Big(\int_0^1e^{\lambda_{in}(1-s)}v_i(s)\dd s,\ \int_0^1e^{\lambda_{in}s}v_i(s)\dd s\Big)\in\C^2,\label{eq:hz}\\
(\Hk v)_o(t)&=\sum_iG_{oi}\,\re\Big[(\mathbf M_i\zeta_i)^{\top}\big(e^{\lambda_{i\cdot}t},\,e^{\lambda_{i\cdot}(1-t)}\big)\Big],\label{eq:hh}\\
\mathcal L(v)&=\sigma\big(\Tk v+\Hk v+Wv+b\big),\label{eq:layer}
\end{align}
with complex residues $R^\pm\in\C^{C\times C\times N}$, mixing matrices $\mathbf M_i\in\C^{2N\times2N}$ and gains $G\in\R^{C\times C}$. Where the operator has a local part, the network adds a single direct feedthrough from the input to the output, $\G_\theta(a)=\mathcal Q\circ\mathcal L_L\circ\cdots\circ\mathcal L_1\circ\mathcal P(a)+Da$ with $D\in\R^{d_u\times d_a}$, the $D$ of a state-space realisation $\mathsf C(s\mathsf I-\mathsf A)^{-1}\mathsf B+D$. Unlike $W$, which acts on the hidden channels inside each layer, $D$ is one linear map that bypasses lifting, layers and projection. It is an input--output skip connection: it adds no expressive power beyond what a skip connection gives, but it makes a linear local map exactly representable and allows the least-squares initialisation of Section~\ref{sec:training}. In two dimensions the kernel is the tensor product $\re\sum_{p,r}R_{oipr}\,b^{(1)}_{ip}(\tau_1)b^{(2)}_{ir}(\tau_2)$ of two-sided exponential bases, with Hankel-type terms along each axis.
\end{definition}

\begin{corollary}[Exact representation]\label{cor:exact}
Every operator of the form \eqref{eq:structure} with $m$ modes is realised exactly by one HNO layer with identity activation, one channel and $N=m$ poles. In particular, under the assumptions of Theorem~\ref{thm:structure}, one layer represents the solution operator of a well-posed linear constant-coefficient ODE boundary-value problem of order $m$ on an interval.
\end{corollary}

\begin{proof}
Take $\lambda_n=\mu_n$ from \eqref{eq:bstates}. The causal branch of \eqref{eq:green} is $R^+e^{\mu\tau}$, the anti-causal branch $-r_ne^{\lambda_n\tau}=-r_ne^{\mu_n|\tau|}$ for $\tau<0$ is $R^-$, \eqref{eq:hz} are the boundary states, $\mathbf M_i$ is $-\mathbf M^{-1}\widetilde{\mathcal B}$, and \eqref{eq:hh} is $\Psi_o$. Because $\kappa$ and $\G f$ are real, taking real parts in \eqref{eq:hk} and \eqref{eq:hh} changes nothing, and conjugate pole pairs need only one representative in the pole bank.
\end{proof}

\begin{remark}[Scope of the exact representation]\label{rem:scope}
Corollary~\ref{cor:exact} uses the two-sided splitting of Theorem~\ref{thm:structure}: a root with $\re\lambda_n>0$ is represented by a decaying exponential anchored at the right end point, $e^{-\lambda_n(1-t)}$, which is anti-causal. This does not carry over to causal initial-value problems with unstable roots. There the impulse response $g$ of Theorem~\ref{thm:causal} contains the growing term $e^{\lambda_nt}$, and a causal layer whose poles all satisfy $\re\lambda<0$ cannot reproduce it exactly; on a finite window it can only approximate it, with constants that grow with the window length. The exact causal representation therefore requires $\re\lambda_n<0$ for all roots, and marginal roots ($\re\lambda_n=0$) are reached only in the limit of Remark~\ref{rem:past}.
\end{remark}

\begin{proposition}[Length-independent bounds]\label{prop:bounded}
Let the layer act on $L^2(0,\ell)$ for an interval of any length $\ell$, and write $\eta_n=-\re\lambda_n>0$. For every channel pair and fixed parameters, $\|\Tk\|\le\sum_n(|R^+_n|+|R^-_n|)/\eta_n$, and $\Hk$ is an operator of rank at most $4N$ per input channel with $\|\Hk\|\le c\,\|G\|\,\|\mathbf M\|\sum_n1/(2\eta_n)$, where $c$ depends only on $N$. Both bounds are independent of $\ell$. They are not uniform over the parameters: they grow without limit as $\eta_n\to0$ or as the residues grow.
\end{proposition}

\begin{proof}
Young's inequality $\|\kappa*v\|_2\le\|\kappa\|_1\|v\|_2$ with $\int_0^\infty|e^{\lambda\tau}|\dd\tau=1/\eta$. $\Hk$ is a finite sum of rank-one operators $\langle\varphi,\cdot\rangle\psi$ whose factors are exponentials $e^{\lambda s}$, $e^{\lambda(\ell-s)}$ with $\|e^{\lambda\cdot}\|^2_{L^2(0,\ell)}=(1-e^{-2\eta\ell})/(2\eta)\le1/(2\eta)$; the real part doubles the number of terms.
\end{proof}

On a bounded interval every integral operator with a bounded kernel is bounded, so the content of Proposition~\ref{prop:bounded} is the independence of $\ell$: a single pole with $\re\lambda>0$, which LNO does not exclude, gives a norm that grows like $e^{\re\lambda\,\ell}/\re\lambda$ with the length of the domain.

\begin{proposition}[FFT evaluation of the non-periodic quadrature]\label{prop:discr}
On a uniform grid $t_j=j/(n-1)$, approximate $\Tk v$ by the trapezoid rule with half weight at the kernel jump $\tau=0$. (i) This discrete sum is a linear (non-circular) discrete convolution, and zero padding to length $2n$ followed by the FFT evaluates it exactly, up to round-off, in $O(C^2n\log n)$ operations. (ii) For $v\in C^2$ the discrete sum approximates the continuous integral with error $O(n^{-2})$ at the grid points. The parameters are independent of $n$, so the same layer acts on any grid.
\end{proposition}

\begin{proof}
On each side of $\tau=0$ the integrand is $C^2$, and the trapezoid rule on each one-sided segment has error $O(n^{-2})$; the half weight at $s=t$ is the endpoint weight of both segments. Zero padding to $2n$ removes the wrap-around of the circular convolution. The kernel is sampled at physical lags, so changing $n$ changes only the quadrature nodes.
\end{proof}

\begin{figure}[pos=htbp]
\centering
\includegraphics[width=0.92\textwidth]{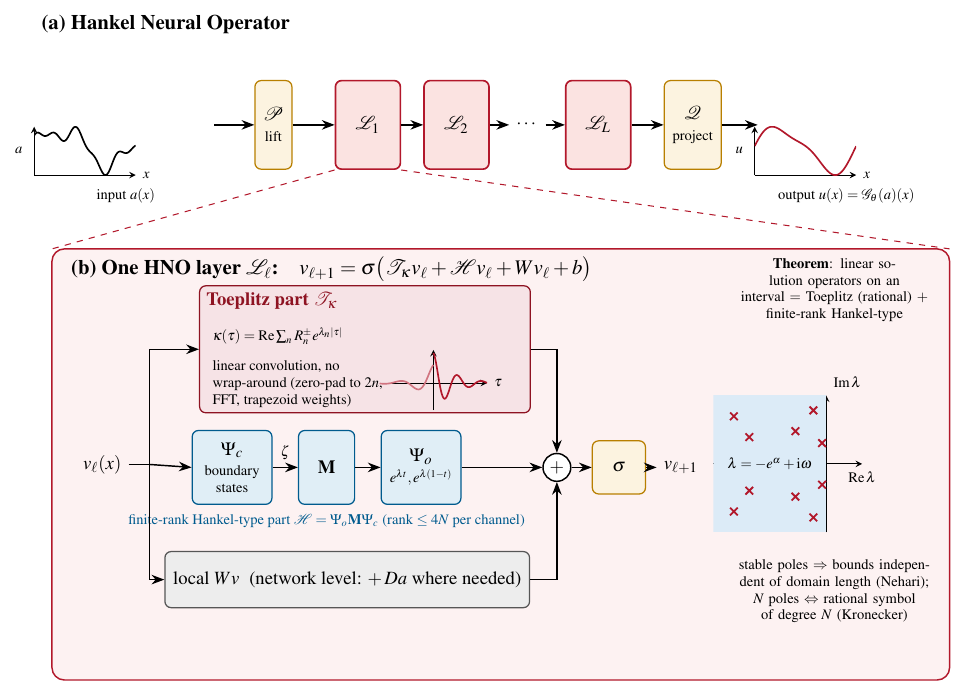}
\caption{Architecture of HNO. (a) Lifting $\mathcal P$, $L$ HNO layers and projection $\mathcal Q$. (b) One layer: rational Toeplitz part $\Tk$ (non-periodic convolution evaluated by the FFT), finite-rank Hankel-type part $\Psi_o\mathbf M\Psi_c$ (boundary and initial-state information) and local part $W$; where needed, an input--output feedthrough $Da$ is added at the network level. All poles lie in the open left half-plane.}
\label{fig:overview}
\end{figure}

In our implementation the FFT evaluation agrees with the direct $O(n^2)$ evaluation of the same discrete sum to $2\times10^{-16}$ (part (i)), and the quadrature error with respect to the integral falls by a factor of four per grid refinement (part (ii)). For the time axis of an initial-value problem (Theorem~\ref{thm:causal}) the anti-causal residues are removed and the Hankel-type part reads only the initial value,
\begin{equation}
(\Hk v)_o(t)=\sum_iG_{oi}\,\re\big[(\mathbf M_iv_i(0))^{\top}e^{\lambda_{i\cdot}t}\big].
\label{eq:hcausal}
\end{equation}
Each layer, and hence the network, is then causal. Fig.~\ref{fig:overview} summarises the architecture.

\subsection{Closed-loop (Lur'e) HNO for dynamics without fading memory}
\label{sec:lure}

\begin{definition}[Fading memory \cite{boyd1985fading}]
A causal time-invariant operator $\G$ has fading memory if there is a decreasing weight $\varpi:\R_+\to(0,1]$ with $\varpi(\tau)\to0$ such that $\G$ is continuous at the current time in the norm $\|f\|_\varpi=\sup_{s\le t}|f(s)|\varpi(t-s)$.
\end{definition}

Boyd and Chua \cite{boyd1985fading} proved that every causal time-invariant operator with fading memory is approximated uniformly, on uniformly bounded inputs, by a finite-dimensional stable linear system followed by a static nonlinear read-out (a Wiener model). The benchmarks are posed on a finite window $[0,T]$, so we state the corresponding restriction of feed-forward operators quantitatively.

\begin{proposition}[Feed-forward HNOs have exponentially fading memory]\label{prop:wiener}
Let $\G_\theta$ be a causal feed-forward HNO (causal Toeplitz--Hankel layers \eqref{eq:hcausal}, no closed loop) with $L$ layers, poles with $\re\lambda\le-\eta_0<0$ and Lipschitz pointwise nonlinearities. There are constants $A_\theta,B_\theta$, independent of $T$, such that for all inputs $f,g$ and $0\le\tau\le t\le T$,
\begin{equation}
|\G_\theta(f)(t)-\G_\theta(g)(t)|\le A_\theta\sup_{s\in[t-\tau,t]}|f(s)-g(s)|+B_\theta\,e^{-\eta_0\tau/L}\sup_{s\in[0,t]}|f(s)-g(s)| .
\label{eq:fading}
\end{equation}
\end{proposition}

\begin{proof}
For one causal layer, split the convolution at lag $\tau$: the part with lags in $[0,\tau]$ is bounded by $\|\kappa\|_{L^1}$ times the recent difference, and the part with lags larger than $\tau$ by $\int_\tau^\infty|\kappa|\le e^{-\eta_0\tau}\sum_n|R_n|/\eta_n$ times the global difference. The Hankel-type term \eqref{eq:hcausal} depends on $v(0)$ through $e^{\lambda t}$ with $|e^{\lambda t}|\le e^{-\eta_0 t}\le e^{-\eta_0\tau}$, and $W$, $D$ are pointwise. Hence \eqref{eq:fading} holds for one layer with exponent $\eta_0\tau$. Composition with Lipschitz maps preserves the form; applying the one-layer estimate $L$ times with window $\tau/L$ in each layer gives the stated exponent.
\end{proof}

Proposition~\ref{prop:wiener} quantifies the limitation. Suppose that changing the input before time $t-\tau$, by at most 1 in the supremum norm, changes the true output at time $t$ by at least $c_0>0$, independently of $\tau$, as for the free oscillation of an undamped oscillator excited early in the window or for the choice of equilibrium of a bistable system. A feed-forward HNO that approximates $\G$ to accuracy $\delta_0<c_0/2$ must then satisfy $B_\theta e^{-\eta_0T/L}\ge c_0-2\delta_0$. So either the slowest pole decays at a rate $\eta_0\lesssim L\log(B_\theta/(c_0-2\delta_0))/T$, which tends to zero as the window grows, or the constant $B_\theta$, which scales like $\sum_n|R_n|/\eta_n$, grows exponentially in $T$. On a fixed window such a model is not excluded, but it lies in a poorly conditioned corner of the parameter space. Proposition~\ref{prop:wiener} is an upper bound on memory. It does not apply to non-causal layers or to the closed loop below, and it does not show that a feed-forward model fails to fit a given finite-window data set. The estimate applies equally to other feed-forward architectures whose layers have exponentially decaying memory. The undamped Duffing oscillator ($u''+u=f-u^3$), the undamped pendulum ($u''+u=f-(\sin u-u)$) and the Lorenz system (a linear part with the quadratic couplings $-xz$ and $xy$) are all of Lur'e form, i.e.\ a linear system in feedback with a static nonlinearity \cite{khalil2002nonlinear}. This suggests closing the HNO loop.

\begin{definition}[Lur'e--HNO]\label{def:lure}
Let $z_{in}(t)\in\C$ be the modal state of the rational kernel, $\dot z_{in}=\lambda_{in}z_{in}+e_i$, with output $y_o=\re\sum_{i,n}R_{oin}z_{in}+(We)_o$. For $e=v$ this is the causal Toeplitz part of Definition~\ref{def:layer}. The Lur'e--HNO closes the loop with a static nonlinearity and uses a linear read-out,
\begin{equation}
e=v-F(y),\qquad F(0)=0,\qquad u=\mathcal Qy ,
\label{eq:lure}
\end{equation}
so that $\G_\theta$ is the solution map of $\dot z=\Lambda z+\mathsf B(v-F(\mathsf Cz))$ with $\Lambda=\operatorname{diag}(\lambda_{in})$ and read-out $\mathsf C$ given by the residues. The open-loop poles are stable, but the closed loop may have marginal or unstable modes, so $\G_\theta$ need not have fading memory.
\end{definition}

\textit{Exact time marching.} The window is split into $J$ blocks of length $\Delta_b$. Inside block $m$ the convolution is evaluated in parallel together with the free response of the carried state; between blocks,
\begin{equation}
z_{m+1}=e^{\lambda\Delta_b}z_m+\int_{\text{block }m}e^{\lambda(\Delta_b-s)}e(s)\dd s .
\label{eq:march}
\end{equation}

\begin{proposition}[Semigroup exactness]\label{prop:semigroup}
For $F\equiv0$ the block-marched Lur'e--HNO equals the full-window causal HNO layer for every $J$.
\end{proposition}

\begin{proof}
For $s\le t_m\le t$, $e^{\lambda(t-s)}=e^{\lambda(t-t_m)}e^{\lambda(t_m-s)}$, so the contribution of all earlier blocks to block $m$ is $e^{\lambda(t-t_m)}z_m$; \eqref{eq:march} is the same identity at $t=t_{m+1}$.
\end{proof}

\begin{proposition}[Picard sweeps]\label{prop:picard}
Inside a block, $e=v-F(\mathcal G_be+\text{free})$, where $\mathcal G_b$ is the block convolution plus the local term $W$, with $\|\mathcal G_b\|_{L^\infty\to L^\infty}\le\|W\|+\sum_n|R_n|(1-e^{-\eta_n\Delta_b})/\eta_n\le\|W\|+\Delta_b\sum_n|R_n|$. If $\Lip(F)\|\mathcal G_b\|<1$, the fixed point is unique and $q$ Picard sweeps converge geometrically. The dynamic part of the bound is $O(\Delta_b)$, so shorter blocks improve the contraction.
\end{proposition}

\begin{proof}
Banach's fixed-point theorem for $e\mapsto v-F(\mathcal G_be+\text{free})$ on $L^\infty$ of the block; the norm of $\mathcal G_b$ is at most $\|W\|$ plus the $L^1$ norm of the kernel on $[0,\Delta_b]$.
\end{proof}

Numerically, the block-marched and full-window layers agree to $10^{-15}$ for $F\equiv0$. We use $q=1$ sweep by default; Table~\ref{tab:ablation} reports $q=2$. The propagation of the linear state between blocks is exact (Proposition~\ref{prop:semigroup}). With a finite number of sweeps, however, the computed map only approximates the continuous closed loop \eqref{eq:lure}, and the approximation is controlled only when the contraction condition of Proposition~\ref{prop:picard} holds, which we do not verify for the trained models. The network that is trained and tested is the discrete $J$-block, $q$-sweep map itself. Algorithm~\ref{alg:lure} gives the forward pass. The first sweep and the output pass are linear in the block inputs, so the convolutions of all blocks are computed in one batched contraction before and after the march. Only the free response, $F$ and the update \eqref{eq:march} are sequential. Independent seeds are trained together as one vectorised ensemble with separate optimiser states.

\begin{algorithm}[htbp]
\caption{Forward pass of the Lur'e--HNO}
\label{alg:lure}
\begin{algorithmic}[1]
\REQUIRE input $v$ split into blocks $v_0,\dots,v_{J-1}$; poles $\lambda$ ($\re\lambda<0$); block convolution $\mathcal G_b$; feedback $F$; sweeps $q$
\STATE $z_0\gets0$; compute $\mathcal G_bv_m$ for all $m$ in one batched contraction
\FOR{$m=0$ \TO $J-1$}
\STATE free$_m\gets$ read-out of $e^{\lambda(t-t_m)}z_m$ on block $m$
\STATE $e_m\gets v_m-F(\mathcal G_bv_m+\text{free}_m)$
\FOR{$k=2$ \TO $q$}
\STATE $e_m\gets v_m-F(\mathcal G_be_m+\text{free}_m)$
\ENDFOR
\STATE $z_{m+1}\gets e^{\lambda\Delta_b}z_m+\int_{\text{block }m}e^{\lambda(\Delta_b-s)}e_m(s)\dd s$ \hfill (exact, Proposition~\ref{prop:semigroup})
\ENDFOR
\STATE $y\gets\mathcal G_b[e_0,\dots,e_{J-1}]+[\text{free}_0,\dots,\text{free}_{J-1}]$ (one batched contraction)
\RETURN $u=\mathcal Qy$
\end{algorithmic}
\end{algorithm}

\subsection{Rational-multiplier HNO on periodic and padded domains}
\label{sec:rational}

\begin{lemma}[Periodisation of an exponential kernel]\label{lem:periodise}
On the unit torus, the periodisation of $\kappa(x)=r\,e^{\lambda x}\one_{x\ge0}$ with $\re\lambda<0$ is $\kappa_{\rm per}(x)=r\,e^{\lambda x}/(1-e^{\lambda})$ on $[0,1)$, and its Fourier coefficients are $\hat\kappa_{\rm per}(k)=r/(2\pi\ii k-\lambda)$. For $\kappa(x)=r\,e^{\lambda|x|}\one_{x<0}$ they are $r/(-2\pi\ii k-\lambda)$.
\end{lemma}

\begin{proof}
$\sum_{j\ge0}e^{\lambda(x+j)}=e^{\lambda x}/(1-e^{\lambda})$ and $\int_0^1e^{(\lambda-2\pi\ii k)x}\dd x=(e^{\lambda}-1)/(\lambda-2\pi\ii k)$; the product of the two factors is $1/(2\pi\ii k-\lambda)$. The anti-causal case follows from $x\mapsto-x$.
\end{proof}

On a torus the Toeplitz part of an HNO layer is therefore the rational Fourier multiplier
\begin{equation}
\hat K_{oi}(\xi)=\sum_{n=1}^NR^+_{oin}\phi_{in}(\xi)+R^-_{oin}\phi_{in}(-\xi),\qquad
\phi_{in}(\xi)=\frac{\eta_{in}}{\ii\xi-\lambda_{in}}=\frac{\eta_{in}}{\eta_{in}+\ii(\xi-\omega_{in})},
\label{eq:rational}
\end{equation}
evaluated at the non-negative physical frequencies $\xi=2\pi k/|\Omega|$, $k\ge0$. The values at negative frequencies follow from conjugate symmetry (Lemma~\ref{lem:hermitian}), so the realised kernel is the real part of the complex rational kernel, as in \eqref{eq:hk}. Each $\phi_{in}$ is a unit Lorentzian of width $\eta_{in}$ centred at $\omega_{in}$, so $|\phi_{in}|\le1$. In $d$ dimensions the multiplier is the tensor product $\sum_{p,r}R_{oipr}\phi_{ip}(\pm\xi_1)\phi_{ir}(\pm\xi_2)$.

\begin{lemma}[Hermitian projection]\label{lem:hermitian}
A real kernel satisfies $\hat K(-\xi)=\overline{\hat K(\xi)}$, so $\hat K(0)\in\R$ and, on an even grid, the Nyquist coefficient is real. When \eqref{eq:rational} is evaluated on $\xi\ge0$ and inverted with a real-output inverse FFT, the kernel is well defined only after the projection $\Pi$: the real part of the DC (and Nyquist) bin in 1D, and $c(k_1)\mapsto\frac12(c(k_1)+\overline{c(-k_1)})$ on the $k_2=0$ (and Nyquist) column in 2D.
\end{lemma}

\begin{proof}
Conjugate symmetry of the Fourier transform of a real function.
\end{proof}

\begin{definition}[Rational-multiplier HNO layer]\label{def:rational}
\begin{equation}
\mathcal L(v)=\sigma\Big(\Fo^{-1}\big[\Pi\big(\hat K(\xi)+\Phi(k)\one_{|k|<k_{\max}}\big)\Fo v\big]+Wv+b\Big),
\label{eq:rlayer}
\end{equation}
with $\hat K$ from \eqref{eq:rational}, free low-mode weights $\Phi$, optional instance normalisation around the integral operator, and the Hermitian projection $\Pi$ of Lemma~\ref{lem:hermitian}. For non-periodic problems the input is zero-padded by a fixed fraction $\varepsilon$ of the domain ($\varepsilon=0.1$). By Lemma~\ref{lem:periodise}, \eqref{eq:rlayer} then realises the linear convolution with the two-sided exponential kernel of Definition~\ref{def:layer} up to wrap-around terms: for a kernel term $r_ne^{\lambda_nx}$ on the unit domain they are bounded pointwise by $|r_n|e^{-\eta_n\varepsilon}/(1-e^{-\eta_n(1+\varepsilon)})$.
\end{definition}

\begin{proposition}[FNO is the case $R^\pm=0$]\label{prop:fno}
With $R^\pm=0$, \eqref{eq:rlayer} is an FNO layer with $k_{\max}$ modes; with $\Phi=0$ it is a pure finite-Hankel-rank layer. The rational part acts at every frequency, has $2N$ residues per channel pair and $N$ poles per channel independent of the grid, and decays as $|\hat K(\xi)|=O(1/|\xi|)$.
\end{proposition}

Lemma~\ref{lem:hermitian} is not a technicality. Without $\Pi$, some FFT back-ends discard the imaginary part of the DC coefficient and others do not. In our first GPU runs this made the Burgers model trained at $n=1024$ fail at $n=8192$ (error 0.24); with $\Pi$ the error is $6.0\times10^{-3}$ at every resolution, and the trained weights are unchanged.

\begin{remark}[Fractional padding]\label{rem:padding}
FNO pads a fixed number of grid points. On a finer grid the padded region is physically narrower, so the learned periodic kernel acts on a different extension of the input, and the operator changes with the resolution. Padding by a fixed fraction of the domain, together with poles in physical units, makes the parameterisation independent of the resolution. Whether the errors are also robust is an empirical question; Table~\ref{tab:transfer} reports it for the tested grids.
\end{remark}

\subsection{Comparison with FNO and LNO}
\label{sec:compare}

\begin{figure}[pos=htbp]
\centering
\includegraphics[width=0.92\textwidth]{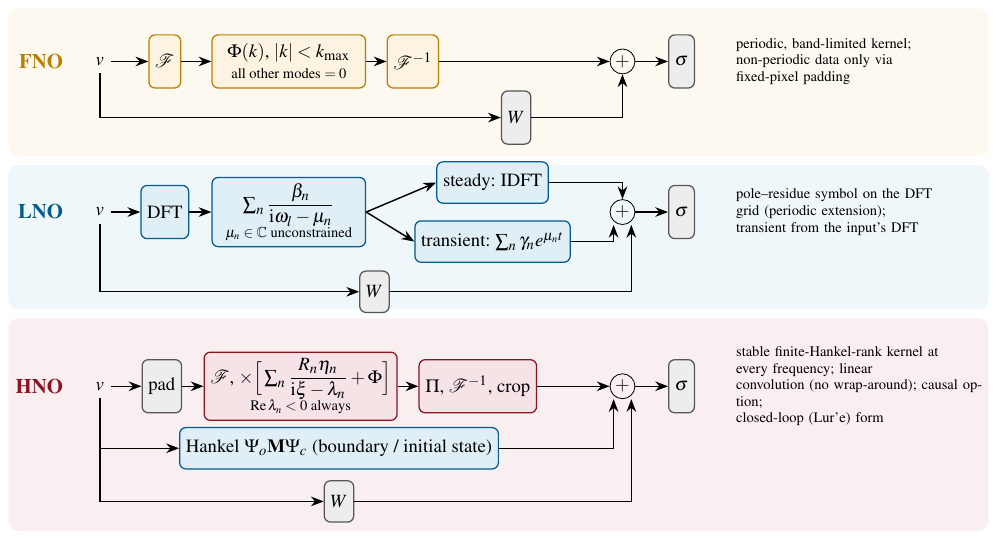}
\caption{One layer of FNO, LNO and HNO. FNO keeps free Fourier modes $|k|<k_{\max}$; LNO evaluates an unconstrained pole--residue symbol on the DFT grid with a transient term; HNO uses stable poles, a non-periodic convolution evaluated by the FFT, a finite-rank Hankel-type term and an optional input--output feedthrough $D$.}
\label{fig:compare}
\end{figure}

Table~\ref{tab:compare} and Fig.~\ref{fig:compare} compare the integral operators. Relative to FNO, HNO represents the Toeplitz and Hankel-type parts of Theorem~\ref{thm:structure} exactly, neither of which is band-limited, and on periodic problems it contains FNO as a special case (Proposition~\ref{prop:fno}). Relative to LNO, the pole--residue idea is shared, but HNO evaluates the non-periodic convolution without wrap-around, obtains the transient (Hankel-type) part from the boundary states that Theorem~\ref{thm:structure} shows to be sufficient, keeps every pole stable, and produces non-fading memory only through the closed loop. In the multi-layer LNO the pole--residue and Fourier branches are separate networks whose outputs are added; in HNO the rational and free parts are two parameterisations of one multiplier inside each layer.

\begin{table}[pos=htbp]
\centering\small
\caption{Integral operators of FNO, LNO and HNO.}
\label{tab:compare}
\begin{tabular}{@{}P{2.8cm}P{3.2cm}P{3.6cm}P{5.5cm}@{}}
\toprule
& FNO & LNO & HNO\\
\midrule
kernel class & band-limited, periodic & pole--residue symbol & stable pole--residue kernel of finite Hankel rank; rational multiplier plus free modes on tori\\
convolution & circular & circular (steady part) & linear (no wrap-around); periodic only where the problem is\\
boundary / initial state & via $W$ and padding & transient term from the DFT of the input & finite-rank Hankel-type term $\Psi_o\mathbf M\Psi_c$ from boundary states\\
pole stability & -- & unconstrained & $\re\lambda<0$; bounds independent of domain length\\
causality on time axes & no & no & yes (Theorem~\ref{thm:causal})\\
local part & $W$ & $W$ & $W$; optional input--output feedthrough $D$\\
non-fading memory & no & no & closed-loop Lur'e--HNO\\
grid change & fixed-pixel padding alters $\G_\theta$ & DFT estimate depends on grid & physical poles, fractional padding\\
cost per layer & $O(C^2n\log n)$ & $O(C^2Nn)$ & $O(C^2n\log n+C^2Nn)$\\
\bottomrule
\end{tabular}
\end{table}

\subsection{Initialisation, training and choice of realisation}
\label{sec:training}

\textit{Pole tiling.} In our experiments a pole frequency moved to the correct value only from within roughly one third of a cycle over the window, so randomly placed frequencies were easily trapped. Poles are therefore initialised to tile the frequency axis: channel $c$ of $C$ uses $\omega=(k+c/C)\Delta\omega$, $k=0,\dots,N-1$, with $\Delta\omega=2\pi f_{\max}/N$ and damping $0.5\Delta\omega$, similar to the starting poles of vector fitting \cite{gustavsen1999rational}. For long memory, the damping of channel $c$ follows a geometric ladder from near-undamped to strongly damped.
\textit{Local-first start.} Where the operator has a local part, $D$ is initialised by least squares and the residues and the projection start at zero, so the network starts as the best local map and adds non-local structure only where it lowers the loss.
\textit{Optimisation.} Adam \cite{kingma2015adam} with weight decay on all parameters except the poles. Lur'e--HNOs use gradient-norm clipping at 1.0.
\textit{Model order.} For the diagonal realisation $\dot z=\Lambda z+\mathsf b e$, $y=\mathsf cz$ of one channel pair of a trained layer, the Gramians are available in closed form, $P_{jk}=-\mathsf b_j\bar{\mathsf b}_k/(\lambda_j+\bar\lambda_k)$ and $Q_{jk}=-\bar{\mathsf c}_j\mathsf c_k/(\bar\lambda_j+\lambda_k)$. The square roots of the eigenvalues of $PQ$ are the Hankel singular values, which bound the error of reducing that linear kernel to $r$ poles (Theorem~\ref{thm:aak}); the effect on the full nonlinear network has to be checked empirically. In this study the number of poles was chosen on validation data; Hankel-singular-value selection is part of the research plan (Section~\ref{sec:limits}).

\begin{figure}[pos=htbp]
\centering
\includegraphics[width=0.92\textwidth]{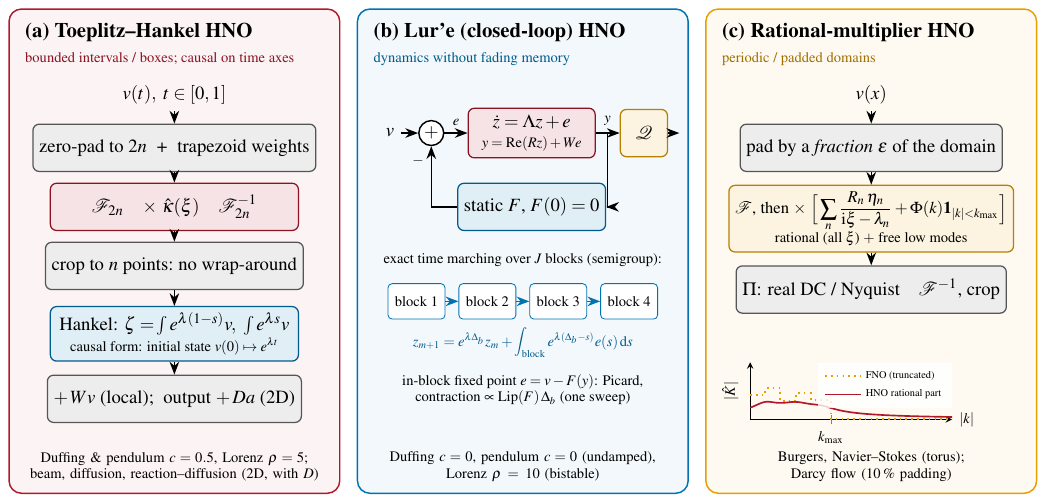}
\caption{The three HNO realisations and the problems on which each is used: (a) Toeplitz--Hankel HNO, (b) closed-loop Lur'e--HNO with block marching (exact linear state propagation), (c) rational-multiplier HNO, which, unlike FNO's truncated multiplier, acts at every frequency.}
\label{fig:instances}
\end{figure}

The three realisations share the pole bank, the kernel and the training protocol. They differ in the domain and in whether the loop is closed, and the choice follows from two properties of the problem (Fig.~\ref{fig:instances}). Fading-memory dynamics use the causal Toeplitz--Hankel HNO (Theorem~\ref{thm:causal}, Proposition~\ref{prop:wiener}); dynamics without fading memory use the Lur'e--HNO; bounded-domain maps with a local part add $D$; periodic PDEs use the rational multiplier; and non-periodic boxes use the rational multiplier with fractional padding. Within a realisation, width, number of poles and depth are chosen on the validation split.

%============================= NUMERICAL STUDY ===============================
\section{Numerical Study}
\label{sec:numerics}

\subsection{Benchmarks and protocol}
\label{sec:setup}

We use the nine problems of the LNO suite \cite{cao2024lno} and the three problems of the FNO suite \cite{li2021fno} (Table~\ref{tab:problems}). Every model is trained with the relative $L^2$ loss \eqref{eq:risk} from input--output data only; no governing equation enters the loss. All reported numbers are test relative $L^2$ errors at the final epoch (mean $\pm$ standard deviation over seeds). Model selection among HNO variants uses only the validation split. $p$ denotes the two-sided Welch $t$-test between HNO and the better baseline. All runs used PyTorch \cite{paszke2019pytorch} on two NVIDIA T4 GPUs (15\,GB usable memory each) on the Kaggle platform, with the jobs distributed over both GPUs.

\textit{LNO suite.} Each problem has 200 training, 50 validation and 130 test samples. The six ODE problems use 2048 points on $t\in[0,20.47]$. Training forcings are $f(t)=A\sin(5t)$ with $A\in[0.05,10]$, while test forcings are $f(t)=Ae^{-0.05t}\sin(5t)$ \cite{cao2024lno}. The test set is therefore an unseen input family, and the benchmark measures extrapolation. The equations are
\begin{equation}
\begin{aligned}
&\text{Duffing:} && u''+cu'+u+u^3=f, && u(0)=u'(0)=0, && c\in\{0,\,0.5\},\\
&\text{pendulum:} && u''+cu'+\sin u=f, && u(0)=u'(0)=0, && c\in\{0,\,0.5\},\\
&\text{Lorenz:} && \dot x=10(y-x),\ \ \dot y=x(\rho-z)-y,\ \ \dot z=xy-\tfrac83z-f, && (x,y,z)(0)=(1,0,0), && \rho\in\{5,\,10\},
\end{aligned}
\label{eq:odes}
\end{equation}
with $\G:f\mapsto u$ and $\G:f\mapsto x$, respectively \cite{kovacic2011duffing,lorenz1963deterministic}. Re-integrating \eqref{eq:odes} reproduces the released Lorenz data to a relative error of $4\times10^{-4}$ ($\rho=5$) and $9\times10^{-4}$ ($\rho=10$). The 2D problems map a source field $f(x,t)$ to $u(x,t)$: the Euler--Bernoulli beam ($50\times50$ grid), diffusion ($50\times50$) and reaction--diffusion ($40\times20$). Training and test sources are again different families \cite{cao2024lno}. Baselines: the official single-layer LNO with its per-problem configuration, and a four-layer FNO with the mode counts of \cite{cao2024lno}. Common protocol: Adam (learning rate $2\times10^{-3}$, weight decay $10^{-4}$), step decay by 0.5 every 100 epochs, 1000 epochs (1200 for the pendulum), batch sizes of \cite{cao2024lno}, 5 seeds.

\textit{FNO suite.} 1000 training, 100 validation and 200 test samples; Adam (learning rate $10^{-3}$, weight decay $10^{-4}$), step decay by 0.5 every 100 epochs, 500 epochs, batch 20, 3 seeds \cite{li2021fno}. The data were regenerated on the GPU with the published FNO generators (same priors, solvers and resolutions), because the original files were not reliably downloadable on the compute platform. Burgers: $\partial_tu+\partial_x(u^2/2)=\nu\partial_{xx}u$ on $\T$, $\nu=10^{-3}$, $\G:u(\cdot,0)\mapsto u(\cdot,1)$, trained at $n=1024$. Darcy: $-\nabla\cdot(a\nabla u)=1$ on $(0,1)^2$ with $u=0$ on the boundary, $a\in\{3,12\}$ a thresholded Gaussian field, $\G:a\mapsto u$, trained at $85^2$. Navier--Stokes: vorticity form on $\T^2$, $\nu=10^{-3}$, forcing $0.1(\sin2\pi(x_1+x_2)+\cos2\pi(x_1+x_2))$, $\G:\omega|_{[0,10]}\mapsto\omega|_{(10,50]}$ at $64^2$, with 10 input and 40 output frames as channels for all models. Baselines: FNO in its official configurations (Darcy with 9-point padding) and the official multi-layer LNO, which is much stronger here than the single-layer LNO (Burgers: 0.20 versus 0.66 in screening). As a capacity control we also train FNO with the parameter count of HNO (FNO$^{+}$).

\begin{table}[pos=htbp]
\centering\small
\caption{Problems, structural property and selected HNO (TH: Toeplitz--Hankel; RM: rational multiplier).}
\label{tab:problems}
\begin{tabular}{@{}rllll r@{}}
\toprule
\# & problem & operator & structure & HNO realisation & params\\
\midrule
1 & Duffing ($c=0$) & $f\mapsto u$ & no fading memory & Lur'e, $2\times8$ poles & 184\\
2 & Duffing ($c=0.5$) & $f\mapsto u$ & fading memory & causal TH, 2 layers & 6\,665\\
3 & Pendulum ($c=0$) & $f\mapsto u$ & no fading memory & Lur'e, $4\times8$ poles & 632\\
4 & Pendulum ($c=0.5$) & $f\mapsto u$ & fading memory & causal TH, 2 layers & 6\,665\\
5 & Lorenz ($\rho=5$) & $f\mapsto x$ & stable equilibrium & causal TH, 1 layer & 6\,553\\
6 & Lorenz ($\rho=10$) & $f\mapsto x$ & bistable & Lur'e, $6\times16$ poles & 2\,423\\
7 & Euler--Bernoulli beam & $f\mapsto u$ & local map & 2D TH $+D$, 1 layer & 40\,274\\
8 & Diffusion & $f\mapsto u$ & local map & 2D TH $+D$, 2 layers & 78\,178\\
9 & Reaction--diffusion & $f\mapsto u$ & local, quadratic & 2D TH $+D$, $8\times8$ poles & 151\,122\\
10 & Burgers & $u_0\mapsto u(\cdot,1)$ & periodic, fronts & RM, $N=4$, 16 free modes & 813\,761\\
11 & Darcy & $a\mapsto u$ & non-periodic, elliptic & RM, $\varepsilon=0.1$ padding & 2\,894\,337\\
12 & Navier--Stokes & $\omega_{[0,10]}\mapsto\omega_{(10,50]}$ & periodic, forced & RM, instance norm & 2\,899\,656\\
\bottomrule
\end{tabular}
\end{table}

\subsection{LNO benchmark suite}
\label{sec:lno-results}

\begin{figure}[pos=htbp]
\centering
\includegraphics[width=0.9\textwidth]{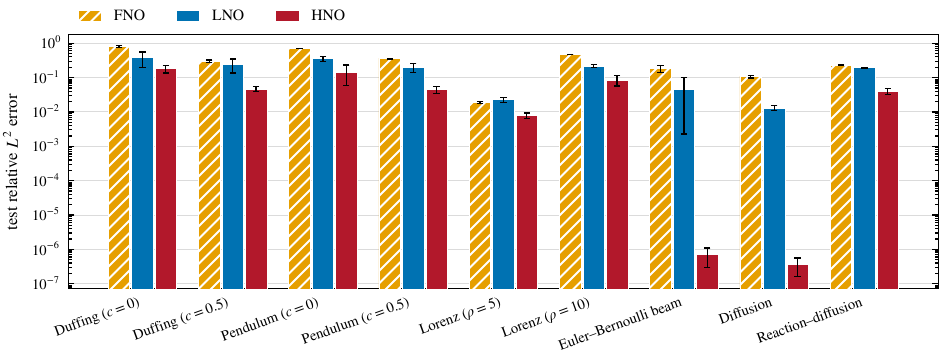}
\caption{LNO suite: test relative $L^2$ error (mean $\pm$ std over 5 seeds, log scale).}
\label{fig:summary}
\end{figure}

Table~\ref{tab:lno} and Fig.~\ref{fig:summary} give the results. HNO has the lowest reported mean error on all nine problems, and its spread over seeds is smaller than LNO's on six of them. The differences are significant at the 5\,\% level on seven problems; the exceptions are Duffing with $c=0$ ($p=0.09$) and the beam ($p=0.16$), where LNO's seed-to-seed variation is large.

\begin{table}[pos=htbp]
\centering\small
\caption{LNO suite: test relative $L^2$ error (5 seeds). $p$: Welch test of HNO against the better baseline.}
\label{tab:lno}
\begin{tabular}{@{}lllll@{}}
\toprule
problem & FNO & LNO & HNO & $p$\\
\midrule
Duffing ($c=0$) & $(7.89\pm0.51)\times10^{-1}$ & $(3.78\pm1.78)\times10^{-1}$ & $\mathbf{(1.80\pm0.44)\times10^{-1}}$ & 0.091\\
Duffing ($c=0.5$) & $(3.03\pm0.22)\times10^{-1}$ & $(2.41\pm1.06)\times10^{-1}$ & $\mathbf{(4.64\pm0.78)\times10^{-2}}$ & 0.021\\
Pendulum ($c=0$) & $(7.02\pm0.15)\times10^{-1}$ & $(3.53\pm0.61)\times10^{-1}$ & $\mathbf{(1.45\pm0.86)\times10^{-1}}$ & 0.005\\
Pendulum ($c=0.5$) & $(3.53\pm0.08)\times10^{-1}$ & $(1.94\pm0.56)\times10^{-1}$ & $\mathbf{(4.43\pm0.97)\times10^{-2}}$ & 0.005\\
Lorenz ($\rho=5$) & $(1.92\pm0.12)\times10^{-2}$ & $(2.25\pm0.33)\times10^{-2}$ & $\mathbf{(7.85\pm1.51)\times10^{-3}}$ & $<10^{-3}$\\
Lorenz ($\rho=10$) & $(4.67\pm0.06)\times10^{-1}$ & $(2.17\pm0.19)\times10^{-1}$ & $\mathbf{(8.59\pm2.92)\times10^{-2}}$ & $<10^{-3}$\\
Euler--Bernoulli beam & $(1.85\pm0.42)\times10^{-1}$ & $(4.53\pm5.27)\times10^{-2}$ & $\mathbf{(6.94\pm3.97)\times10^{-7}}$ & 0.16\\
Diffusion & $(1.06\pm0.09)\times10^{-1}$ & $(1.31\pm0.19)\times10^{-2}$ & $\mathbf{(3.56\pm1.98)\times10^{-7}}$ & $<10^{-3}$\\
Reaction--diffusion & $(2.32\pm0.11)\times10^{-1}$ & $(1.94\pm0.06)\times10^{-1}$ & $\mathbf{(4.07\pm0.85)\times10^{-2}}$ & $<10^{-3}$\\
\bottomrule
\end{tabular}
\end{table}

\begin{figure}[pos=htbp]
\centering
\includegraphics[width=\textwidth]{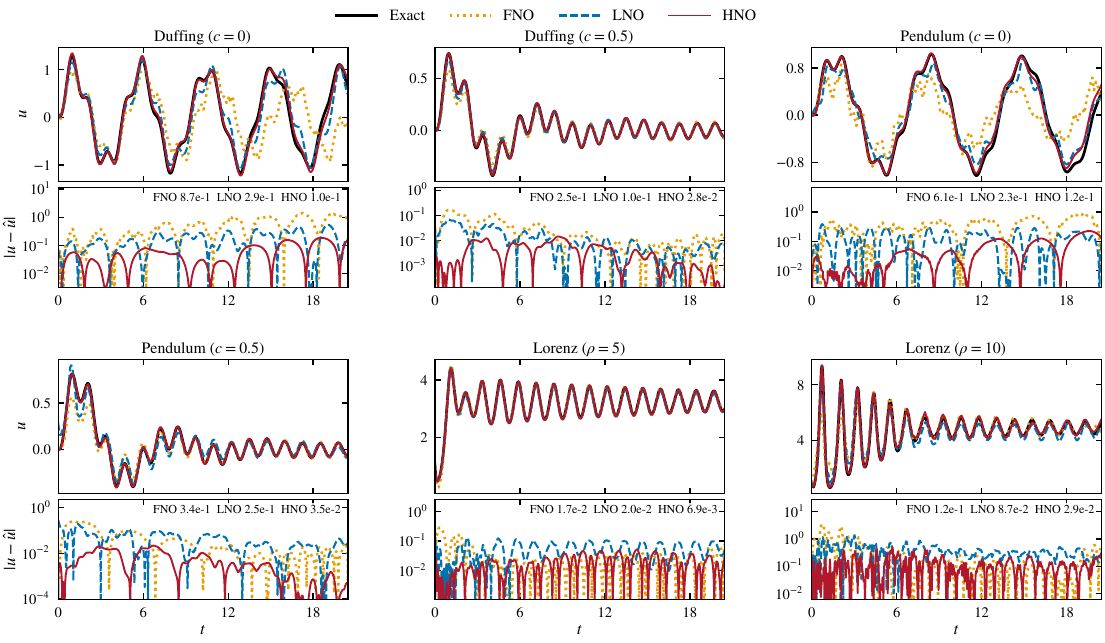}
\caption{LNO suite, ODE problems: exact response and predictions for a median test sample (top) and pointwise error (bottom, log scale); per-sample errors in the legend boxes.}
\label{fig:ode}
\end{figure}

\textit{Damped oscillators ($c=0.5$).} These operators are causal with fading memory, the setting of Theorem~\ref{thm:causal} and Proposition~\ref{prop:wiener}. The causal two-layer HNO reduces LNO's error 5.2 times (Duffing) and 4.4 times (pendulum) with 6\,665 parameters, compared with 8.4 million for FNO. Fig.~\ref{fig:ode} shows that HNO follows the initial transient, where FNO and LNO make their largest errors.

\textit{Undamped oscillators ($c=0$).} Here the late response is the free oscillation launched by the early forcing, which never decays, so the operator has no fading memory. FNO, whose kernel has no state, fails (0.79 and 0.70). The feed-forward causal HNO also stalls at 0.50--0.59, as expected from Proposition~\ref{prop:wiener}. The Lur'e--HNO, with 184 and 632 parameters, halves LNO's error (0.180 versus 0.378 and 0.145 versus 0.353).

\textit{Lorenz system.} For $\rho=5$ all trajectories relax to the equilibrium $C^+=(\sqrt{8(\rho-1)/3},\sqrt{8(\rho-1)/3},\rho-1)$ \cite{strogatz2015nonlinear}, the memory fades, and the causal HNO is 2.4 times more accurate than FNO. For $\rho=10$ the forcing amplitude selects one of the two equilibria: in the released data every sample with $A\gtrsim7.15$ ends on the negative branch. This is a bistable latch, i.e.\ memory that does not fade. For the seed whose predictions are stored, FNO and the feed-forward HNO (error 0.41) always predict the positive branch, which is correct for 79\,\% of the test samples, and LNO picks the correct branch for 95\,\%. The Lur'e--HNO picks it for all 130 samples and reaches 0.086, 2.5 times below LNO. Measured over the four quarters of the window, its error stays between 0.10 and 0.13, while that of the feed-forward HNO grows from 0.34 to 0.92 (seed 0; see also Fig.~\ref{fig:ode}). This growth is consistent with Proposition~\ref{prop:wiener}.

\begin{figure}[pos=htbp]
\centering
\includegraphics[width=\textwidth]{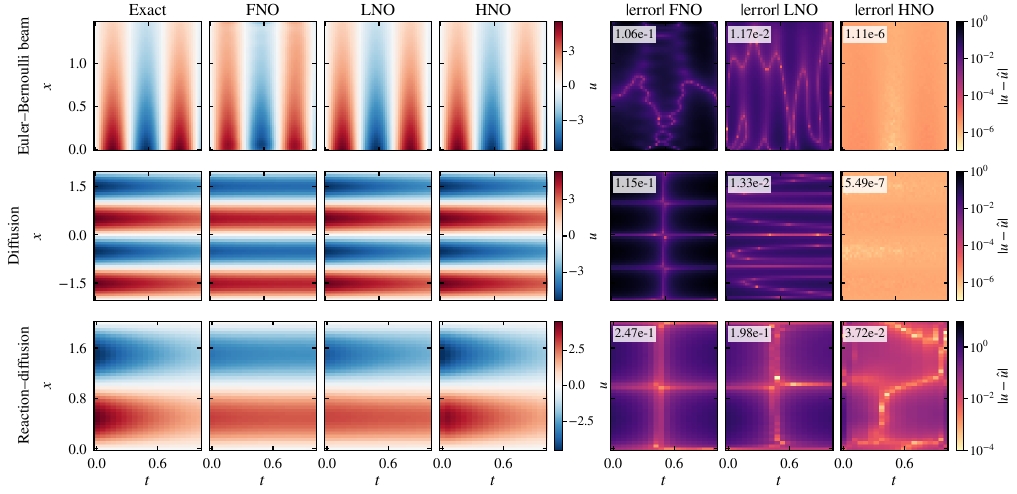}
\caption{LNO suite, 2D problems: exact solution, predictions and pointwise errors (log colour bars) for a median test sample.}
\label{fig:pde}
\end{figure}

\textit{Two-dimensional problems.} On the released data the source and the solution satisfy exact pointwise relations on both splits, to a relative residual of $10^{-16}$: $f=-99u$ (beam), $f=(1-\pi^2)u$ (diffusion) and $f=(1-\pi^2)u+u^2$ (reaction--diffusion). The reason is that the manufactured solutions are single separable modes, for which every derivative in the PDE is a multiple of $u$. The learnable operator is therefore local, which in state-space terms is the feedthrough $D$. The baselines as published have no input--output skip of this kind, although one could be added to them; part of the gain on these problems is due to this skip and its least-squares initialisation rather than to the rational kernel. HNO with $D$ reaches $10^{-7}$ on the two linear problems and 0.041 on reaction--diffusion, where the quadratic term must be built by the nonlinear path (Fig.~\ref{fig:pde}). We state this property of the benchmark openly: these three problems test whether an architecture contains the local operator, not non-local operator learning.

\begin{figure}[pos=htbp]
\centering
\includegraphics[width=\textwidth]{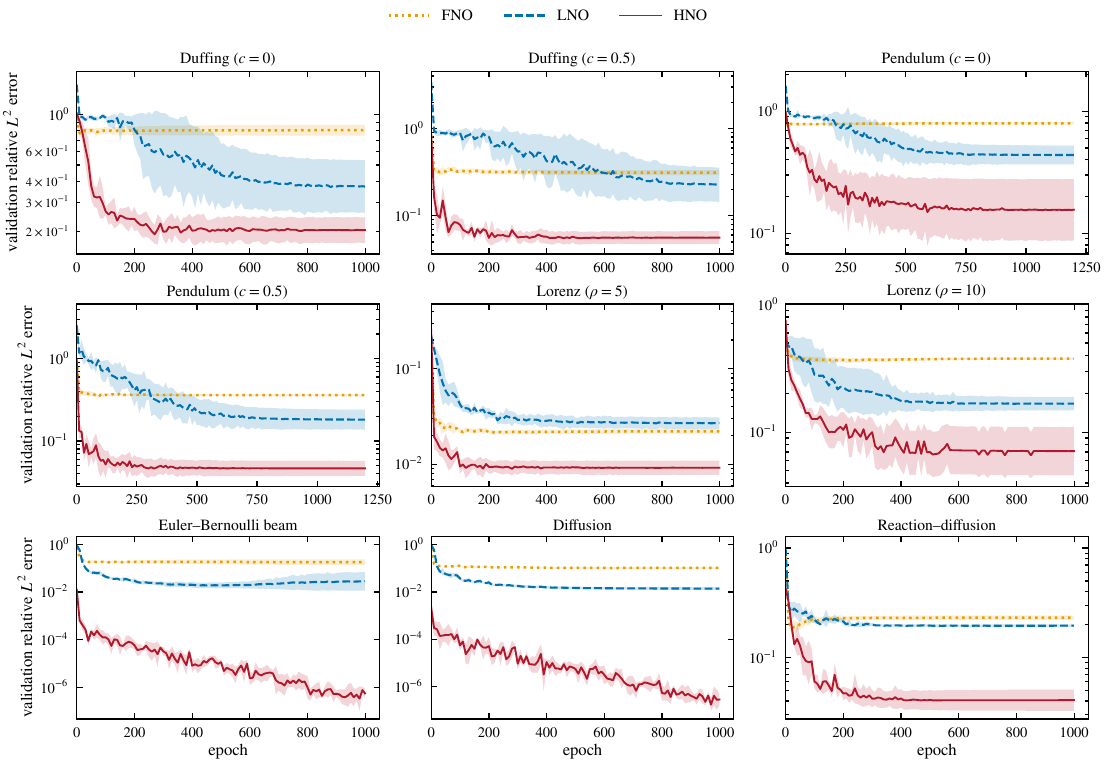}
\caption{LNO suite: validation relative $L^2$ error versus epoch for FNO, LNO and the selected HNO of Table~\ref{tab:problems}. Lines: geometric mean over 5 seeds; bands: $\pm1$ standard deviation of $\log_{10}$ of the error.}
\label{fig:curves}
\end{figure}

\textit{Training behaviour.} Fig.~\ref{fig:curves} shows the validation relative $L^2$ error during training. On all nine problems HNO falls below the final error of LNO within the first 100 epochs. On the ODE problems its curve is flat after 400--600 epochs, whereas on the beam and diffusion problems it keeps decreasing until the last epoch, as expected for an operator that the architecture represents exactly (the local map $D$). The wide LNO bands on Duffing with $c=0$ and on the beam reflect the seed-to-seed spread reported in Table~\ref{tab:lno}. FNO reaches a plateau within about 50 epochs on the undamped and bistable problems, at an error that does not improve afterwards, consistent with a kernel that carries no state across the window.

\subsection{FNO benchmark suite}
\label{sec:fno-results}

\begin{table}[pos=htbp]
\centering\small
\caption{FNO suite: test relative $L^2$ error (3 seeds). FNO$^{+}$: FNO with HNO's parameter count ($^{\dagger}$1 seed).}
\label{tab:fno}
\begin{tabular}{@{}llllll@{}}
\toprule
problem & FNO & LNO & HNO & $p$ & FNO$^{+}$\\
\midrule
Burgers ($n=1024$) & $(7.02\pm0.10)\times10^{-3}$ & $(7.43\pm0.80)\times10^{-2}$ & $\mathbf{(6.00\pm0.10)\times10^{-3}}$ & $<10^{-3}$ & $6.67\times10^{-3\,\dagger}$\\
Darcy ($85^2$) & $(7.88\pm0.06)\times10^{-3}$ & $(4.69\pm0.12)\times10^{-2}$ & $\mathbf{(5.52\pm0.05)\times10^{-3}}$ & $<10^{-3}$ & $(7.55\pm0.08)\times10^{-3}$\\
Navier--Stokes ($64^2$) & $(2.45\pm0.01)\times10^{-2}$ & $(3.79\pm0.12)\times10^{-2}$ & $\mathbf{(2.24\pm0.02)\times10^{-2}}$ & $<10^{-3}$ & $(2.27\pm0.03)\times10^{-2}$\\
\bottomrule
\end{tabular}
\end{table}

\begin{figure}[pos=htbp]
\centering
\includegraphics[width=\textwidth]{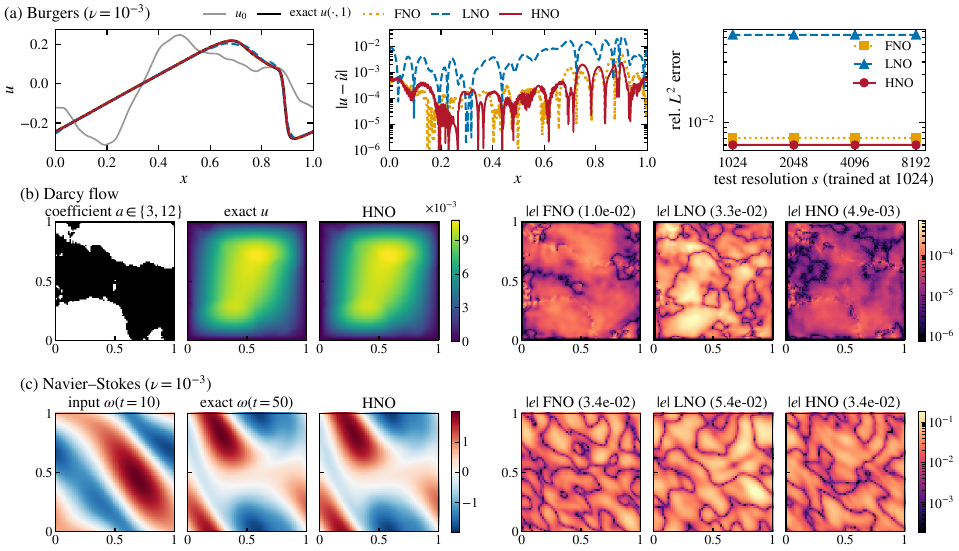}
\caption{FNO suite. (a) Burgers: initial condition, solution and predictions, pointwise error, and zero-shot resolution transfer. (b) Darcy and (c) Navier--Stokes: input, exact solution, HNO prediction and pointwise errors (shared log colour bar; sample error in brackets).}
\label{fig:fno}
\end{figure}

\begin{figure}[pos=htbp]
\centering
\includegraphics[width=\textwidth]{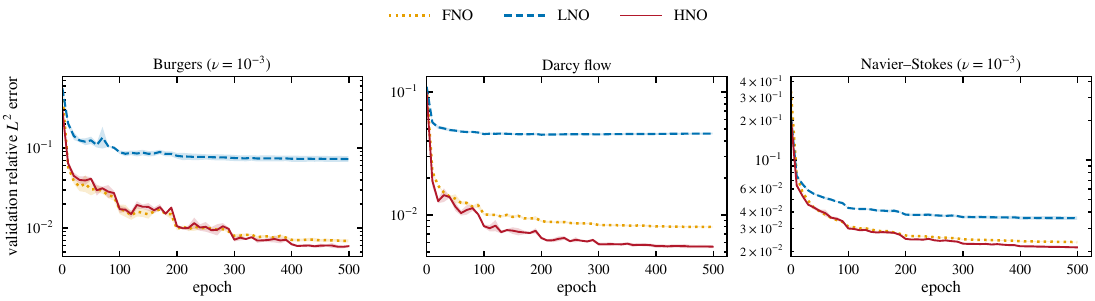}
\caption{FNO suite: validation relative $L^2$ error versus epoch (geometric mean over 3 seeds; bands as in Fig.~\ref{fig:curves}).}
\label{fig:fnocurves}
\end{figure}

Table~\ref{tab:fno} and Figs.~\ref{fig:fno}--\ref{fig:fnocurves} show the results. HNO has the lowest mean error on all three problems. The gains over FNO are 15\,\% on Burgers, 30\,\% on Darcy and 9\,\% on Navier--Stokes, smaller than on the LNO suite, as expected: FNO was designed for these problems. Matching FNO's parameter count to HNO's does not close the gap on Burgers or Darcy ($p<10^{-4}$ for Darcy). On Navier--Stokes it does: FNO$^{+}$ and HNO are statistically indistinguishable ($p=0.26$), so the gain there is mainly capacity. The parameter-matched Burgers control was run with one seed only, because of the compute budget, and is indicative. The validation curves (Fig.~\ref{fig:fnocurves}) show that HNO and FNO converge at a similar rate on Burgers and Navier--Stokes, and that HNO separates from FNO early on Darcy flow, where fractional padding and the rational multiplier act on the non-periodic box.

\subsection{Zero-shot resolution transfer}
\label{sec:transfer}

Table~\ref{tab:transfer} evaluates trained models on other grids without retraining. On Darcy flow, HNO's error stays between $5.4\times10^{-3}$ and $5.7\times10^{-3}$ from $85^2$ to $421^2$, while FNO's grows from $7.9\times10^{-3}$ to $6.7\times10^{-2}$, which is the effect of fixed-pixel padding described in Remark~\ref{rem:padding}. LNO's error is grid independent but about 8.5 times higher than HNO's, and LNO ran out of memory at $421^2$. On Burgers all three models are constant from $n=1024$ to $8192$; HNO stays at $6.0\times10^{-3}$ once the Hermitian projection is applied. On the LNO suite (inputs subsampled $2\times$ and $4\times$), HNO changes by less than 13\,\% on eight problems, whereas FNO's error on diffusion grows fivefold. The exception is reaction--diffusion, where HNO's error grows five-fold at $4\times$ coarser inputs (Section~\ref{sec:limits}). The LNO-suite tests only sub-sample the training grid; finer grids were not tested there.

\begin{table}[pos=htbp]
\centering\small
\caption{Zero-shot resolution transfer (test relative $L^2$ error, mean over seeds). LNO suite: $1$, $\frac12$, $\frac14$ of the training resolution.}
\label{tab:transfer}
\resizebox{\textwidth}{!}{%
\begin{tabular}{@{}l ccc ccc ccc@{}}
\toprule
& \multicolumn{3}{c}{FNO} & \multicolumn{3}{c}{LNO} & \multicolumn{3}{c}{HNO}\\
\cmidrule(lr){2-4}\cmidrule(lr){5-7}\cmidrule(lr){8-10}
LNO suite & $1$ & $\tfrac12$ & $\tfrac14$ & $1$ & $\tfrac12$ & $\tfrac14$ & $1$ & $\tfrac12$ & $\tfrac14$\\
\midrule
Duffing ($c=0$) & 7.89e-1 & 7.90e-1 & 7.90e-1 & 3.78e-1 & 3.89e-1 & 4.13e-1 & \textbf{1.80e-1} & \textbf{1.77e-1} & \textbf{1.74e-1}\\
Duffing ($c=0.5$) & 3.03e-1 & 3.03e-1 & 3.03e-1 & 2.41e-1 & 2.43e-1 & 2.47e-1 & \textbf{4.64e-2} & \textbf{4.55e-2} & \textbf{4.42e-2}\\
Pendulum ($c=0$) & 7.02e-1 & 7.02e-1 & 7.03e-1 & 3.53e-1 & 3.58e-1 & 3.69e-1 & \textbf{1.45e-1} & \textbf{1.50e-1} & \textbf{1.63e-1}\\
Pendulum ($c=0.5$) & 3.53e-1 & 3.53e-1 & 3.54e-1 & 1.94e-1 & 1.96e-1 & 2.02e-1 & \textbf{4.43e-2} & \textbf{4.47e-2} & \textbf{4.59e-2}\\
Lorenz ($\rho=5$) & 1.92e-2 & 1.95e-2 & 2.08e-2 & 2.25e-2 & 2.20e-2 & 2.08e-2 & \textbf{7.85e-3} & \textbf{7.82e-3} & \textbf{7.89e-3}\\
Lorenz ($\rho=10$) & 4.67e-1 & 4.68e-1 & 4.69e-1 & 2.17e-1 & 2.22e-1 & 2.31e-1 & \textbf{8.59e-2} & \textbf{8.60e-2} & \textbf{8.68e-2}\\
Beam & 1.85e-1 & 1.85e-1 & 1.84e-1 & 4.53e-2 & 4.84e-2 & 5.48e-2 & \textbf{6.94e-7} & \textbf{6.87e-7} & \textbf{6.89e-7}\\
Diffusion & 1.06e-1 & 5.66e-1 & 5.75e-1 & 1.31e-2 & 1.74e-2 & 2.64e-2 & \textbf{3.56e-7} & \textbf{3.55e-7} & \textbf{3.59e-7}\\
Reaction--diffusion & 2.32e-1 & 4.56e-1 & 4.57e-1 & 1.94e-1 & 2.07e-1 & 2.96e-1 & \textbf{4.07e-2} & \textbf{1.10e-1} & \textbf{2.17e-1}\\
\bottomrule
\end{tabular}}

\medskip
\begin{tabular}{@{}lllll@{}}
\toprule
FNO suite & test grid & FNO & LNO & HNO\\
\midrule
Darcy (trained at $85^2$) & $85^2$ & 7.88e-3 & 4.69e-2 & \textbf{5.52e-3}\\
 & $141^2$ & 3.20e-2 & 4.70e-2 & \textbf{5.40e-3}\\
 & $211^2$ & 4.81e-2 & 4.71e-2 & \textbf{5.48e-3}\\
 & $421^2$ & 6.70e-2 & -- (memory) & \textbf{5.70e-3}\\
Burgers (trained at $1024$) & $1024$--$8192$ & 7.02e-3 (constant) & 7.43e-2 (constant) & \textbf{6.00e-3 (constant)}\\
\bottomrule
\end{tabular}
\end{table}

\subsection{Ablation study}
\label{sec:ablations}

Each row of Table~\ref{tab:ablation} changes one component of a trained HNO and keeps everything else, including the protocol, fixed. The results tie each component to the theory.

\begin{table}[pos=htbp]
\centering\scriptsize
\caption{Ablations: each row changes one component of the reference HNO (mean over 3--5 seeds; $^\dagger$1 seed). Entries are reference $\to$ changed.}
\label{tab:ablation}
\begin{tabular}{@{}P{1.9cm}P{3.6cm}P{4.2cm}P{5.5cm}@{}}
\toprule
group & change & problems & test rel.\ $L^2$: reference $\to$ changed\\
\midrule
Lur'e--HNO & no feedback ($F=0$) & Duffing ($c=0$); pendulum ($c=0$); Lorenz ($\rho=10$) & $0.180\to0.869$;\ \ $0.145\to0.585$;\ \ $0.086\to0.529$\\
 & MLP instead of linear read-out & same three & $0.180\to0.340$;\ \ $0.145\to0.204$;\ \ $0.086\to0.122$\\
 & over-sized ($8\times16$ modes) & same three & $0.180\to0.697$;\ \ $0.145\to0.377$;\ \ $0.086\to\mathbf{0.049}$\\
 & blocks $J=32$; $J=128$; $J{=}64,q{=}2$ & Duffing ($c=0$) & $0.180\to0.161$;\ \ $0.146$;\ \ $0.143$\\
 & blocks $J=32$; $J=64$ & Lorenz ($\rho=10$) & $0.086\to0.191$;\ \ $0.089$\\
\midrule
Toeplitz--Hankel & non-causal kernel & Duffing, pendulum ($c=0.5$); Lorenz ($\rho=5$) & $0.097\to0.309$;\ \ $0.096\to0.369$;\ \ $0.018\to0.066$\\
 & non-causal kernel & Duffing, pendulum ($c=0$); Lorenz ($\rho=10$) & $0.441\to0.809$;\ \ $0.442\to0.597$;\ \ $0.415\to0.490$\\
 & no Hankel-type (state) term & Duffing, pendulum ($c=0.5$) & $0.095\to0.095$;\ \ $0.071\to0.074$\\
 & no Hankel-type (state) term & Lorenz ($\rho=5$); Lorenz ($\rho=10$); Duffing ($c=0$) & $0.0079\to0.0086$;\ \ $0.408\to0.413$;\ \ $0.585\to0.662$\\
 & unconstrained poles & six ODE problems & change within $\pm3\,\%$\\
 & random pole initialisation & Lorenz ($\rho=5$); Duffing ($c=0$) & $0.0079\to0.0157$;\ \ $0.585\to\mathbf{0.417}$\\
 & no feedthrough $D$ & beam; diffusion; reaction--diffusion & $\sci{6.9}{-7}\to0.034$;\ \ $\sci{8.9}{-7}\to0.017$;\ \ $0.054\to0.071$\\
\midrule
Rational multiplier & no free modes & Burgers ($N{=}8$); Darcy & $\sci{6.57}{-3}\to\sci{9.30}{-3}$;\ \ $\sci{5.52}{-3}\to\sci{5.71}{-3}$\\
 & rational part band-limited & Burgers ($N{=}8$) & $\sci{6.57}{-3}\to\sci{6.90}{-3\,\dagger}$\\
 & $N=8$ instead of $N=4$ poles & Burgers & $\sci{6.00}{-3}\to\sci{6.57}{-3}$\\
 & instance normalisation switched & Burgers (on); Darcy (on) & $\sci{6.00}{-3}\to\sci{4.34}{-2}$;\ \ $\sci{5.52}{-3}\to\sci{9.49}{-3}$\\
 & instance normalisation switched & Navier--Stokes (off) & $\sci{2.24}{-2}\to\sci{2.37}{-2}$\\
\midrule
Loss & ODE residual added & Duffing, pendulum ($c=0.5$) & $0.055\to\mathbf{0.046}$;\ \ $0.056\to\mathbf{0.037}$\\
\bottomrule
\end{tabular}

\smallskip\raggedright\footnotesize Bold: the change improves on the reference (discussed in Section~\ref{sec:limits}). References are the HNO variant that the change modifies.
\end{table}

(i) \textit{Feedback is what supplies non-fading memory.} Removing $F$ from the Lur'e--HNO returns the errors to the level of a linear model (0.87, 0.59, 0.53). Replacing the linear read-out by an MLP is also worse (0.34, 0.20, 0.12) and increases the seed spread: once the loop supplies the memory, the linear read-out of Definition~\ref{def:lure} suffices, and extra nonlinearity after the loop mainly overfits. (ii) \textit{Causality is the right bias on time axes.} The non-causal kernel is worse on all six ODE problems, by factors of 1.2 to 3.8, although it has more parameters. (iii) \textit{The Hankel-type term matters only when the initial state is non-zero.} By Theorem~\ref{thm:causal} the true operator has no Hankel-type term when the initial data vanish, which is the case for the Duffing and pendulum problems; removing the term from the layers changes the damped results by less than 5\,\%. It helps for Lorenz $\rho=5$, whose initial state is $(1,0,0)$, by 9\,\%. (iv) \textit{Stability is a guarantee, not an accuracy device.} With unconstrained poles the errors change by at most 3\,\% on the six ODE problems, so the constraint costs no accuracy while it removes the possibility of growing modes. (v) \textit{The local part.} Removing $D$ (and with it the least-squares initialisation) costs four to five orders of magnitude on the linear 2D problems, in line with the pointwise structure of the data. This ablation isolates the skip connection, not the rational kernel. (vi) \textit{Rational and free modes complement each other.} On Burgers the rational part alone is worse than with free low modes, and restricting the rational part to the free-mode band costs 5\,\%, so the tail at high frequencies contributes. Instance normalisation helps on Navier--Stokes, where the vorticity grows under forcing, and hurts on Burgers and Darcy.

\subsection{Computational cost}
\label{sec:cost}

Table~\ref{tab:cost} lists parameters, training time per epoch and inference time for the whole test set on one T4 GPU. On the FNO suite HNO is 1.7--2.7 times slower than FNO per epoch and 1.6--3.9 times faster than the multi-layer LNO. On the LNO suite the feed-forward HNOs are as fast as FNO with three orders of magnitude fewer parameters. The Lur'e--HNOs are the slowest models, because the block march is sequential, but they use only 184--2\,423 parameters, and training one model takes about ten minutes.

\begin{table}[pos=htbp]
\centering\small
\caption{Cost: parameters, training s/epoch and inference ms for the test set (one T4).}
\label{tab:cost}

\begin{tabular}{@{}l rrr rrr rrr@{}}
\toprule
& \multicolumn{3}{c}{FNO} & \multicolumn{3}{c}{LNO} & \multicolumn{3}{c}{HNO}\\
\cmidrule(lr){2-4}\cmidrule(lr){5-7}\cmidrule(lr){8-10}
problem & params & s/ep & ms & params & s/ep & ms & params & s/ep & ms\\
\midrule
Duffing ($c=0$) & 8\,405\,473 & 0.32 & 44 & 1\,821 & 0.09 & 10 & 184 & 0.38 & 46\\
Duffing ($c=0.5$) & 8\,405\,473 & 0.32 & 44 & 1\,821 & 0.09 & 9 & 6\,665 & 0.16 & 14\\
Pendulum ($c=0$) & 8\,405\,473 & 0.26 & 49 & 2\,077 & 0.07 & 10 & 632 & 0.35 & 52\\
Pendulum ($c=0.5$) & 8\,405\,473 & 0.26 & 49 & 1\,309 & 0.06 & 10 & 6\,665 & 0.11 & 17\\
Lorenz ($\rho=5$) & 8\,405\,473 & 0.31 & 49 & 1\,821 & 0.09 & 11 & 6\,553 & 0.10 & 11\\
Lorenz ($\rho=10$) & 8\,405\,473 & 0.31 & 44 & 6\,173 & 0.32 & 15 & 2\,423 & 0.63 & 51\\
Beam & 15\,360\,769 & 0.58 & 111 & 14\,929 & 0.32 & 53 & 40\,274 & 0.30 & 66\\
Diffusion & 34\,956\,033 & 0.70 & 102 & 14\,929 & 0.32 & 53 & 78\,178 & 0.47 & 97\\
Reaction--diffusion & 7\,217\,665 & 0.13 & 21 & 119\,537 & 0.69 & 92 & 151\,122 & 0.10 & 20\\
\midrule
Burgers & 549\,569 & 0.72 & 40 & 275\,266 & 7.61 & 250 & 813\,761 & 1.97 & 101\\
Darcy & 2\,368\,001 & 5.84 & 406 & 644\,802 & 16.34 & 714 & 2\,894\,337 & 9.94 & 664\\
Navier--Stokes & 2\,373\,320 & 2.38 & 155 & 655\,152 & 9.49 & 429 & 2\,899\,656 & 5.00 & 305\\
\bottomrule
\end{tabular}
\end{table}

%============================= DISCUSSION ===============================
\section{Discussion}
\label{sec:discussion}

\subsection{When HNO is the better choice}

The experiments support a simple rule. HNO helps most when the solution operator has structure that a periodic kernel cannot express: initial and boundary information, causality, long memory, or a local part. On the LNO suite, where all of these occur, it lowers the error of the better baseline by factors between 2.1 and 5.2 on the ODEs, and by several orders of magnitude where the operator is local. On periodic or well-padded problems designed for FNO, the gains are moderate (9--30\,\%), and on Navier--Stokes they are matched by a larger FNO. HNO's norm bounds and parameters do not depend on the domain length or the grid, and its parameters (poles, residues, Hankel singular values) can be interpreted in system-theoretic terms. The realisation is chosen from two structural properties, fading memory and periodicity, before training; only its size is tuned on validation data.

\subsection{Limitations and a research plan}
\label{sec:limits}

Table~\ref{tab:limits} lists the limitations we observed, their cause and how we intend to address them.

\begin{table}[pos=htbp]
\centering\small
\caption{Limitations of HNO, their cause, and planned remedies.}
\label{tab:limits}
\begin{tabular}{@{}P{4.3cm}P{5.0cm}P{6.1cm}@{}}
\toprule
limitation & cause & planned remedy\\
\midrule
Lur'e--HNO is the slowest model (2--5$\times$ LNO per epoch) & block march is sequential & parallel associative scan for the linear part \cite{orvieto2023resurrecting}; fixed-point solvers with implicit gradients\\
Lur'e size is problem dependent (over-sized: worse on oscillators, better on Lorenz $\rho=10$) & state dimension of the true system unknown & choose the order from Hankel singular values (Theorem~\ref{thm:aak}) and prune poles during training\\
Two LNO-suite gains not significant ($p=0.09$, $0.16$) & large seed spread of LNO, 5 seeds & more seeds and test sets with several input families\\
Navier--Stokes gain explained by capacity & FNO already suited to periodic, smooth fields & closed-loop space--time HNO for long rollouts; non-periodic flow benchmarks\\
Reaction--diffusion degrades on coarse inputs (0.041 $\to$ 0.217 at $\frac14$) & quadratic local term evaluated on coarse samples & resolution-aware treatment of the nonlinear local path (e.g.\ anti-aliased evaluation \cite{bartolucci2023representation})\\
Feed-forward HNO prefers random poles on undamped problems & tiled poles fit fading memory, not marginal modes & initialise from data by vector fitting or the eigensystem realisation algorithm \cite{gustavsen1999rational,juang1985eigensystem}\\
Rational part alone weaker on fronts (Burgers) & $O(1/|\xi|)$ tail cannot form shocks with few poles & adaptive pole placement near the front's spectrum; multi-resolution poles\\
Benchmarks are narrow (one-parameter forcings, pointwise 2D data, regenerated FNO data); no non-local boundary-value problem with a known Green's function & inherited from the published suites & a non-local boundary-value problem with a known Green's function, with an ablation of the Hankel-type term; heat and wave equations with random initial data; irregular domains\\
Part of the 2D LNO-suite gain is due to the input--output feedthrough $D$ & baselines were run as published, without this skip and without fractional padding & FNO with the same fractional padding, parameter budget and feedthrough, to isolate the effect of the rational kernel\\
\bottomrule
\end{tabular}
\end{table}

Two further points concern scope rather than accuracy. First, HNO is purely data driven, like FNO and LNO. Adding the ODE residual to the loss improves the damped problems (pendulum $c=0.5$: 0.037 versus 0.044 for the selected data-only model), so a physics-informed HNO is a natural extension, but we exclude it from the comparison because the baselines do not use the equation. Second, the structural theorem is proved for linear, constant-coefficient problems on an interval; its use on boxes (tensor products) and for nonlinear problems is justified empirically. Nonlinear operators enter through composition with pointwise maps and, for non-fading memory, through the Lur'e loop. Extending the analysis to variable coefficients, where the Toeplitz part becomes a pseudo-differential operator with a rational symbol in local coordinates, is the main theoretical task ahead.

%============================= CONCLUSION ===============================
\section{Conclusion}
\label{sec:conclusion}

We proposed to build neural operators from the Toeplitz plus finite-rank Hankel-type structure of solution operators. The Hankel neural operator has norm bounds independent of the domain length and grid-independent parameters by construction, represents linear constant-coefficient solution operators on intervals exactly, extends to dynamics without fading memory through a Lur'e loop, and contains FNO as a special case on periodic domains. On all twelve problems of the LNO and FNO benchmark suites, trained from data alone under the original protocols, it has the lowest reported mean test error; the differences on Duffing ($c=0$) and the beam, and the Navier--Stokes comparison with a parameter-matched FNO, are not statistically significant. The gains are largest where the theory predicts them (transients, undamped and bistable dynamics, bounded domains, local operators) and moderate on the periodic problems for which FNO was designed. The ablations connect each component to a statement of the theory, and the remaining weaknesses, mainly the cost of the closed loop and the choice of model order, have concrete remedies. We therefore see HNO as a principled alternative to Fourier and Laplace layers when the solution operator lives on a bounded domain or carries memory, while its advantages on other problem classes remain to be established.

\section*{Declaration of Generative AI and AI-assisted Technologies in the Writing Process}

During the preparation of this manuscript, the authors used Claude (Anthropic) for assistance with programming, data analysis and drafting of the text. After using this tool, the authors carefully reviewed, revised, and validated the content. The authors take full responsibility for the scientific integrity and content of the published article.

\section*{Author Declaration}
The authors have no conflicts of interest to disclose.

\section*{Data Availability}
The LNO benchmark data are publicly available from the repository of Cao et al.\ \cite{cao2024lno}, \url{https://github.com/qianyingcao/Laplace-Neural-Operator}. The FNO benchmark data were regenerated with the published generators of Li et al.\ \cite{li2021fno}; the generation code (\texttt{src/fno\_data.py}: priors, solvers, resolutions and random seeds) and all reference results are publicly available at \hnorepo{}.

\section*{Code Availability}
The source code, the Kaggle notebooks (run on two NVIDIA T4 GPUs) and all reference results are publicly available at \hnorepo{}. The configurations of every model are in \texttt{src/bench.py} and \texttt{src/bench\_fno.py}. The seeds are 0--4 for the LNO suite and 0--2 for the FNO suite. The only single-seed runs are the parameter-matched FNO on Burgers and the band-limited rational ablation on Burgers, marked in Tables~\ref{tab:fno} and~\ref{tab:ablation}. Per-seed results, training curves and seed-0 predictions are included in the repository.

\end{document}